\documentclass[journal]{IEEEtran}
\usepackage{amssymb}
\usepackage{amsfonts}
\usepackage{amsmath}
\usepackage{algorithm}
\usepackage{algorithmic}
\usepackage{cite}
\usepackage{stfloats}
\usepackage{amsmath}
\usepackage{graphicx}
\usepackage{amsfonts,amsmath,amssymb}
\usepackage{url}
\usepackage{bm}
\usepackage{bbm}
\usepackage{subfigure}
\usepackage{algorithmic}
\usepackage{algorithm}
\usepackage{color}

\newtheorem{proposition}{\textbf{Proposition}}
\newtheorem{lemma}{\textbf{Lemma}}
\newtheorem{remark}{\textbf{Remark}}

\usepackage{amssymb}
\usepackage{}
\usepackage{amsfonts}
\usepackage{amsmath}
\usepackage{algorithm}
\usepackage{algorithmic}
\usepackage{cite}

\usepackage{graphicx,subfigure,amsmath,amssymb,cite}
\usepackage{float}
\normalsize
\ifCLASSINFOpdf
\else
\fi
\begin{document}
\title{UAV-Enabled Confidential Data Collection in Wireless Networks}

%\author{Xiaobo Zhou, \IEEEmembership{Student Member, IEEE,} Shihao Yan, \IEEEmembership{Member, IEEE,} Min Li, \IEEEmembership{Member, IEEE,} \\Feng Shu, \IEEEmembership{Member, IEEE,} and Jun Li, \IEEEmembership{Senior Member, IEEE}

\author{Xiaobo Zhou, \IEEEmembership{Student Member, IEEE,} Shihao Yan, \IEEEmembership{Member, IEEE,} Min Li, \IEEEmembership{Member, IEEE,} \\Jun Li, \IEEEmembership{Senior Member, IEEE,} and Feng Shu, \IEEEmembership{Member, IEEE}

\thanks{ X. Zhou is with the Laboratory of Functional Materials and Device for Informatics, Fuyang Normal University, Fuyang, 236037, China, and also with the School of Nanjing
University of Science and Technology, Nanjing, 210094, China. (e-mail: zxb@njust.edu.cn).}
\thanks{S. Yan is with the School of Engineering, Macquarie University, Sydney, NSW 2109, Australia (e-mail: shihao.yan@mq.edu.au).}
\thanks{M. Li is with the College of Information Science and Electronic Engineering, Zhejiang University, Hangzhou 310027, China (e-mail: min.li@zju.edu.cn).}
\thanks{J. Li and F. Shu are with the School of Electronic and Optical Engineering, Nanjing University of Science and Technology, Nanjing, 210094, China (e-mails: \{jun.li,shufeng\}@njust.edu.cn).}
}% <-this % stops a space

\maketitle

\begin{abstract}
\boldmath
This work, for the first time, considers confidential data collection in the context of unmanned aerial vehicle (UAV) wireless networks, where the scheduled ground sensor node (SN) intends to transmit confidential information to the UAV without being intercepted by other unscheduled ground SNs. Specifically, a full-duplex (FD) UAV
collects data from each scheduled SN on the ground and generates artificial noise (AN) to prevent the scheduled SN's confidential information from being wiretapped by other unscheduled SNs. We first derive the reliability outage probability (ROP) and secrecy outage probability (SOP) of a considered fixed-rate transmission, based on which we
formulate an optimization problem that maximizes the minimum average secrecy rate (ASR) subject to some specific constraints. We then transform the formulated optimization problem into a convex problem with the aid of first-order restrictive approximation technique and penalty method.
The resultant problem is a generalized nonlinear convex programming (GNCP) and solving it directly still leads to a high complexity, which motivates us to further approximate this problem as a second-order cone program (SOCP) in order to reduce the computational complexity. Finally, we develop an iteration procedure based on penalty successive convex approximation (P-SCA) algorithm to pursue the solution to the formulated optimization problem.
Our examination shows that the developed joint design achieves a significant performance gain compared to a benchmark scheme.

\end{abstract}

\IEEEpeerreviewmaketitle

\section{Introduction}

%Recently, unmanned aerial vehicles (UAVs) have been widely utilized in agricultural production, such as pesticide spraying, crop growth monitoring, and agricultural data collection. Especially, UAV acts as a mobile data collector
%
%
%For example, UAVs can be deployed as mobile information disseminators or data collectors to assist the emerging Internet of Things (IoT) applications (e.g., \cite{Zhan2018Energy,Lyu2016Cyclical}).
%
%We note that UAV can sufficiently close to each ground sensor node (SN) to enhance the channel quality from the SN to the UAV and decrease the energy consumption of each SN when the UAV is used as a mobile data collector to gather the sensing data from wireless sensor networks (WSNs) in remote areas.

Recently, unmanned aerial vehicles (UAVs) have been widely utilized in wireless communication networks, due to their on-demand deployment, low cost, controllable mobility and high probability of line-of-sight (LoS) air-to-ground link~\cite{Zhou2019Covert}. In general, UAVs mainly serve as mobile base stations, relays,  information disseminators, and data collectors in wireless networks to assist various applications~\cite{Zeng2016Wireless}.
Specifically, UAVs can be used as mobile base stations to increase the coverage area or capacity of the conventional terrestrial wireless networks (e.g., \cite{Zhaonan2019Joint,JointWu2018,UnmanHWANG2018}).
For example, UAV-mounted base station can be employed to recover communication service after the ground communication infrastructure being damaged in natural disasters.
In addition, UAVs can also be employed as mobile relays to provide reliable wireless connections between two or more wireless devices, between which the direct communication links are severely blocked due to large-bodied obstacles (e.g., \cite{Zhang2018Joint,UAV2018Xiao}).
Compared to the traditional static relays on the ground, UAV relays can significantly improve the communication performance, since its optimal deployment location can be dynamically adjusted according to the time-varying communication channels.
Furthermore, UAVs can be deployed as mobile information disseminators or data collectors to assist the emerging Internet of Things (IoT) applications (e.g., \cite{Zhan2018Energy,Lyu2016Cyclical}). For instance, UAVs can move sufficiently close to each ground sensor node (SN) to enhance the channel quality from the SN to the UAVs in wireless sensor networks (WSNs), which are deployed in remote areas.

Due to the inherent broadcast nature of wireless channels, crucial concerns on the wireless communication security are emerging\cite{Chen2017Survey,Chen2018Exploiting,Shu2016Robust,Chen2019Physical}. In UAV networks, it becomes easier for an eavesdropper to intercept the confidential messages transmitted by the UAV, due to the high probability of the existence of the LoS channel between the UAV and the ground eavesdropper, which poses new security challenges in the context of UAV wireless networks\cite{Wu2019Safeguarding}.
%In the context of UAV networks, wireless communication security is of increasing concerns since the strong LoS air-to-ground communication link also poses more stringent security challenge to the UAV wireless communication networks \cite{Wu2019Safeguarding}.
Against this background, several recent works were devoted to addressing the wireless communication security in the context of the UAV networks from the perspective of physical layer security (e.g., \cite{Zhang2019Securing,Tang2019Secrecy,Zhou2018Improving,Zhou2019UAV,Cai2018Dual,Cheng2019UAV,Wang2017Improving,Lei2019Safeguarding,Liu2019Safeguarding,Cui2018Robust}).
In \cite{Zhang2019Securing}, the authors jointly optimized the UAV's trajectory and transmit power  to effectively enhance the channel quality of the legitimate communication link and degrade the channel quality of the eavesdropping link. Meanwhile, utilizing a UAV as a mobile friendly jammer was considered in \cite{Tang2019Secrecy,Zhou2018Improving} to improve the communication security of wireless communication networks, where the UAV jammer aimed to create AN to confuse the eavesdropper. Along this direction, the communication security of the wireless network consisting of two cooperative UAVs and multiple users was examined in \cite{Cai2018Dual,Zhou2019UAV}, where one UAV was deployed to transmit confidential messages to ground users, while the other UAV generated AN to create interference to the eavesdropper. The secure communication in UAV relay networks was considered in \cite{Cheng2019UAV,Wang2017Improving}, where a UAV was utilized as a mobile relay to improve the communication security performance by adjusting its location dynamically. Furthermore, the secrecy rate of a cellular-connected UAV IoT network with spatially distributed eavesdroppers was investigated in \cite{Lei2019Safeguarding}.
In addition to randomly distributed eavesdroppers, a full-duplex (FD) active eavesdropper in the context of UAV networks was considered in \cite{Liu2019Safeguarding}, where the eavesdropper performed both malicious jamming and eavesdropping simultaneously.
Most recently, the authors of \cite{Cui2018Robust} considered a practical scenario, where the UAV only knew each eavesdropper's imperfect location information and the estimation error on the locations of eavesdroppers was assumed within an uncertain circular region. In this scenario, the robust UAV trajectory and transmit power were jointly designed to maximize the average secrecy rate (ASR) for a worse-case scenario.

WSNs are usually deployed in different application scenarios to collect various data, such as in earthquake monitoring, soil moisture monitoring, and wildlife tracking~\cite{Gong2018Flight}. However, in many cases it is difficult to collect the data from sensors in WSNs, since many sensors may be located in remote areas without communication coverage.
Against this background, utilizing a UAV as a data collector is
highly desirable in remote WNSs due to its remarkably advantages in terms of on-demand deployment and high mobility. For example, recent works \cite{Zhan2018Energy} and \cite{Zhan2019Energy} showed that the UAV data collector can sequentially visit each SN and can move sufficiently close to the scheduled SN for enhancing the quality of the communication link from the ground SN to the UAV. In addition, a flight time minimization problem in the context of UAV data collection network was studied in \cite{Gong2018Flight}. However, the wireless communication security of such UAV data collection networks was completely overlooked in the literature, which is a critical issue in some application scenarios. For example, the scheduled SN may prefer to keep its transmitted information confidential from other unscheduled SNs (e.g., a spy intends to transmit the stolen information to his base without being wiretapped by others).

Against the aforementioned background, in this work we address the confidential data collection with the aid of a UAV from the perspective of physical layer security. Specifically, in order to prevent the confidential information transmitted by the scheduled SN from being intercepted by other unscheduled SNs, a FD UAV generates AN to interfere with other SNs when it gathers critical information from the scheduled SN. Our goal is to maximize the minimum ASR among all SNs on the ground via jointly optimizing the UAV's trajectory and AN transmit power as well as the transmission rates and SN scheduling strategy, which is a new design framework that jointly considers the secrecy outage probability (SOP) and the reliability outage probability (ROP) constraints in UAV networks. The main contributions of this work are summarized as below.

\begin{itemize}
\item For the first time, we consider the UAV-enabled confidential data collection from the perspective of physical layer security. The transmission in the considered system is always subject to the reliability and security outages due to the fact that the UAV suffers from the self-interference and the instantaneous channel state information (CSI) from the scheduled SN to the unscheduled SNs is unavailable. As such, in order to facilitate solving the optimal design problem, we first derive an analytical expression for the ROP, which determines the codeword rate based on a given maximum allowable ROP.  We then derive the SOP expression for each unscheduled SNs, which enables us to determine the SOP constraint analytically in our considered system model.

    \item We formulate an optimization problem to determine the UAV trajectory, AN transmit power, transmission rates, and the SN scheduling in order to maximize the minimum ASR by considering the fairness among the $K$ ground SNs. The formulated optimization problem is a mixed-integer non-convex problem, which is hard to solve directly. Thus, we develop an iterative algorithm based on the penalty successive convex approximation (P-SCA) technique to pursue a suboptimal solution to this problem. To this end, we first convert the problem into a continuous optimization problem, then we construct a penalty function that violates the binary constraint, and finally we apply the first-order restrictive approximation method to transform the initial optimization problem into a convex problem.

     \item The resultant optimization problem in each iteration can be categorized as a generalized nonlinear convex programming (GNCP), since it involves a general exponential cone constraint. We note that solving the convex exponential cone often leads to a high computational complexity. Thus, we develop a novel method to transform the GNCP problem into a standard second-order cone program (SOCP), which is of a lower complexity. Numerical results show that the UAV's trajectory has a significant impact on the max-min ASR in the considered system. Our examination also shows that the UAV's trajectory achieved by our developed P-SCA scheme always shrinks inward relative to the region determined by all the ground SNs.
\end{itemize}

The reminder of this work is organized as follows. In Section II, we present the considered system model together with the adopted assumptions. In Section III, we first formulate the optimal design problem and then derive the analytical expressions for ROP and SOP. In Section IV, we develop a P-SCA scheme to tackle the formulated optimization problem in order to jointly design the UAV's trajectory, AN transmit power, transmission rates, and the SN scheduling. Numerical results are presented in Section V to examine the performance of the developed scheme and Section V presents our conclusion remarks.

\begin{figure}[!ht]
  \centering
  % Requires \usepackage{graphicx}
  \includegraphics[width=2.8in]{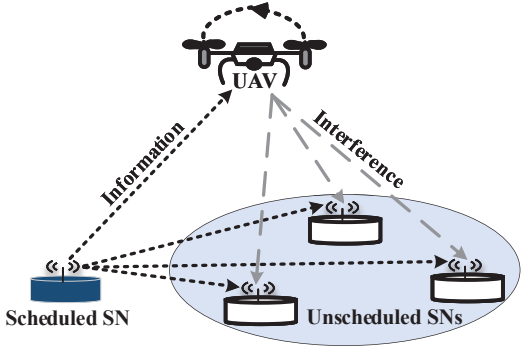}\\
  \caption{UAV-enabled Confidential Data Collection in Wireless Sensor Networks.}\label{Sys_model}
\end{figure}

\section{System Model}

\subsection{Considered Scenario and Adopted Assumptions}

As shown in Fig.~\ref{Sys_model}, in this work we consider a UAV communication network, where a UAV acts as a data collector to gather information from $K$ ground SNs and each SN wants to transmit its confidential information to the UAV without being wiretapped on by other SNs.
We assume that at most one SN is scheduled for communication with the UAV at each time instant.
In order to prevent the
confidential information transmitted by the scheduled SN from being intercepted by other unscheduled SNs, we consider a FD UAV, which is equipped with a receive antenna and a transmit antenna, for secure data collection. Specifically, when the UAV gathers the confidential data from the scheduled $k$-th SN, the UAV also simultaneously generates AN to interfere with other SNs.
The UAV's flight period and flight altitude are set to $T$ and $H$, respectively.
The horizontal coordinates of the $k$-th SN and the UAV are denoted as $\mathbf{w}_k\in\mathbb{R}^{2\times 1}$ ($k\in\mathcal{K}\triangleq\{1,2,\cdots, K\}$) and $\mathbf{q}_u(t)\in\mathbb{R}^{2\times 1}$, where $0\leq t\leq T$. In general, the UAV flies with a limited speed, and its flying speed constraint can be expressed as $\mathbf{\dot{q}}_u(t)\leq V_{\max}$, $0\leq t\leq T$, where $\mathbf{\dot{q}}_u(t)$ denotes the time-derivative of $\mathbf{q}_u(t)$ and $V_{\max}$ is the UAV's maximum flying speed. We note that the UAV's flying speed constraint is continuous with respect to the flying time $t$, which implies an infinite number of speed constraints. To overcome this problem, we divide $T$ into $N$ time slots, i.e., $T=N\delta_t$, where $\delta_t$ denotes the duration of each time slot. Then, at the $n$-th time slot, the UAV's horizontal coordinate is given by $\mathbf{q}_u[n]\in\mathbb{R}^{2\times 1}$ ($n\in \mathcal{N}\triangleq\{1,2\cdots,N\}$).
Following the above clarifications, the mobility constraints of the UAV are given by
\begin{subequations}\label{Mob}
\begin{align}
\mathbf{q}_u[1]&=\mathbf{q}_u[N],\label{Moba}\\
\|\mathbf{q}_u[n+1]-\mathbf{q}_u[n]\|&\leq V_{\max}\delta_t,~n\in \mathcal{N}\setminus\{N\},\label{Mobb}
\end{align}
\end{subequations}
where \eqref{Moba} implies that the UAV has to return to the initial location by the end of the last time slot, while \eqref{Mobb} denotes the UAV's maximum flying distance within each time slot.

Let $P_u[n]$ denote the UAV's AN transmit power at the $n$-th time slot, the peak transmit power constraint is given by
\begin{align}\label{P_u}
 0 \leq P_u[n]\leq P^u_{\max},~\forall n,
\end{align}
where $P^u_{\max}$ denotes the maximum transmit power of the UAV. We assume that at most one SN is scheduled at each time slot, and $\alpha_k[n]\in\{0,1\}$ denotes the SN scheduling indicator, where $\alpha_k[n]=1$ indicates that the $k$-th SN is scheduled for transmission at the $n$-th time slot. As such, the SN scheduling constraint can be expressed as
\begin{align}\label{schedule}
\sum_{k=1}^K \alpha_k[n]\leq 1,\forall n, ~~\alpha_k[n]\in\{0,1\},\forall k, n.
\end{align}
\subsection{Channel Model and Received Signals}
Considering that the air-to-ground and ground-to-air channels are dominated by the LoS \cite{Zhou2019UAV}, the channel from the $k$-th SN to the UAV or the channel from the UAV to the $k$-th SN is given by
\begin{align}\label{channel}
h_{k,u}[n]=h_{u,k}[n]=\sqrt{\frac{\beta_0}{\|\mathbf{q}_u[n]-\mathbf{w}_k\|^2+H^2}},~\forall k, n,
\end{align}
where $\beta_0$ denotes the power gain at a reference
distance $1$ meter (m). The channel from the $k$-th scheduled SN to the $m$-th unscheduled SN and the self-interference channel of UAV are denoted by $g_{k,m}[n]$ ($m\in \mathcal{K}\setminus \{k\}$) and $g_{u,u}[n]$, respectively. We note that $g_{k,m}[n]$ and $g_{u,u}[n]$ are assumed to be subject to quasi-static Rayleigh fading, i.e., $g_{k,m}[n]$ and $g_{u,u}[n]$ follow $\mathcal{CN}(0,\lambda_{k,m})$ and $\mathcal{CN}(0,\lambda_{u,u})$, respectively.
We assume that the scheduled SN only knows the channel distribution information (CDI) between itself and other unscheduled SNs, while the exact instantaneous channel state information (CSI) is unavailable.
%We also assume that each unscheduled SN possesses the full knowledge of instantaneous CSI between it and the unscheduled SN, which is the worse case for the secure communication in the context of the considered scenario.
%Since the randomness of the channel between the ground SNs,
 Considering that the SNs intend to deliver the collected information to the UAV, we assume that the UAV knows the location information of each SN and thus knows the corresponding CSI.

 When the $k$-th SN is scheduled at the $n$-th time slot, the received signal at the UAV is given by
\begin{align}\label{y_u}
y_u[n]=\sqrt{P^k_s[n]}h_{k,u}[n]s_k+\sqrt{\rho P_u[n]}g_{u,u}[n]s_u+z_u,
\end{align}
where $P^k_s[n]$ is the transmit power of the $k$-th scheduled SN, $0\leq\rho\leq 1$  denotes the self-interference cancellation coefficient, and $z_u$ is the Gaussian noise at the UAV with mean $0$ and variance $\sigma_u^2$. In addition, $s_k$ and $s_u$ are confidential signal and AN signal, respectively, which satisfy $\mathbb{E}[|s_k|^2]=1$ and $\mathbb{E}[|s_u|^2]=1$, respectively.

As per \eqref{y_u}, the channel capacity from the $k$-th scheduled SN to the UAV at the $n$-th time slot is given by
\begin{align}\label{Cap_u}
C_{k,u}[n]=\log_2\left(1+\frac{P^k_s[n]|h_{k,u}[n]|^2}{\rho P_u[n]|g_{u,u}[n]|^2+\sigma_u^2}\right).
\end{align}

When the $k$-th SN is scheduled at the $n$-th time slot, the received signal at the $m$-th SN, $m\in \mathcal{K}\setminus \{k\}$, is given by
\begin{align}\label{y_m}
y_m[n]=\sqrt{P^k_s[n]}g_{k,m}[n]s_k+\sqrt{P_u[n]}h_{u,m}[n]s_u+z_m,
\end{align}
where $z_m$ is the Gaussian noise at the $m$-th unscheduled SN with mean $0$ and variance $\sigma_m^2$. The channel capacity from the $k$-th scheduled SN to the $m$-th unscheduled SN can be expressed as
\begin{align}\label{Cap_m}
C_{k,m}[n]=\log_2\left(1+\frac{P^k_s[n]|g_{k,m}[n]|^2}{ P_u[n]|h_{u,m}[n]|^2+\sigma_m^2}\right).
\end{align}

%In addition, we assume that each unscheduled SN knows the channel from itself to the UAV, which is the worse case for the secure communication in the context of the considered scenario.

\subsection{Fixed-Rate Transmission and Outage Probabilities}

In order to achieve secure transmission, the scheduled SN adopts a wiretap code to transmit information to the UAV, in which two rates, i.e., the codeword rate $R_{k,u}[n]$ and the redundancy rate $R_{k,e}[n]$, have to be determined \cite{Zhou2011Rethinking,Yan2018Three,Yan2016Artificial,Yan2015Optimization}. In this work, we consider a fixed-rate transmission from the scheduled SN to the UAV, where $R_{k,u}[n]$ and $R_{k,e}[n]$ are predetermined and to be optimized.

The transmission from the $k$-th SN to the UAV may occur outage since the UAV suffers from the self-interference. As such, when the $k$-th SN is scheduled, the ROP (i.e., reliability outage probability) from the $k$-th SN to the UAV can be expressed as
\begin{align}\label{ROP}
p_{k}^{ro}[n]=\mathrm{Pr}(R_{k,u}[n]>C_{k,u}[n]),
\end{align}
where we recall that $R_{k,u}[n]$ is the codeword rate used for the transmission from SN $k$ to the UAV.

In our considered scenario, perfect secrecy cannot be guaranteed, since the scheduled SN only has the CDI of unscheduled SNs. As such, when the $k$-th SN is scheduled at the $n$-th time slot, the SOP (i.e., secrecy outage probability) is given by
\begin{align}\label{SOP}
p_{k}^{so}[n]=\mathrm{Pr}\left(R_{k,e}[n]<\max_{m\in \mathcal{K}\setminus \{k\}} C_{k,m}[n]\right),
\end{align}
where we recall that $R_{k,e}[n]$ denotes the redundancy rate used to confuse the eavesdroppers (i.e., other unscheduled SNs).

%We note that the adopted SOP, i.e., $p_{k}^{so}[n]$, is different from the existing secrecy outage formulation  (i.e., the probability that the instantaneous secrecy rate is less than a specific value), since the latter includes not only ROP but also SOP. As such, the existing secrecy outage formulation does not distinguish between reliability and the security\cite{Zhou2011Rethinking,Yan2018Three,Yan2016Artificial,Yan2015Optimization}.

In the following section, we first derive exact analytic expressions for SOP and ROP. Then, we jointly design the UAV's trajectory and AN transmit power as well as the code rates and the SN scheduling to maximize the minimum ASR among all SNs subject to some specific constraints.

\section{Optimization Problem Formulation}

For ease of presentation, we define $\mathbf{Q}=\{\mathbf{q}_u[n],\forall n\}$, $\mathbf{A}=\{\alpha_k[n],\forall k, n\}$, $\mathbf{P}_{\mathrm{U}}=\{P_u[n],\forall n\}$,
$\mathbf{R}_{\mathrm{U}}=\{R_{k,u}[n],\forall k, n\}$,
and $\mathbf{R}_{\mathrm{E}}=\{R_{k,e}[n],\forall k, n\}$,
where $\mathbf{q}_u[n]$ is the UAV trajectory, $\alpha_k[n]$ is the scheduling variable, $P_u[n]$ is the UAV's AN transmit power,
$R_{k,u}[n]$ is the transmission rate from SN $k$ to UAV,
and $R_{k,e}[n]$ is the cost
of securing the message transmission of SN $k$ against eavesdropping.
In order to ensure that the UAV can serve each SN and guarantee the fairness among all SNs, our design aim is to maximize the minimum ASR among all SNs by jointly designing the UAV's trajectory $\mathbf{Q}$, the SN scheduling $\mathbf{A}$, the AN transmit power $\mathbf{P}_{\mathrm{U}}$, the transmission rate $\mathbf{R}_{\mathrm{U}}$, and the redundancy rate $\mathbf{R}_{\mathrm{E}}$. The formulated optimization problem is given by
\begin{subequations}\label{PF0}
\begin{align}
&(\mathbf{P1}):\max_{\substack{\mathbf{Q},\mathbf{A},\mathbf{R}_{\mathrm{U}}\\
\mathbf{P}_{\mathrm{U}},\mathbf{R}_{\mathrm{E}}}}\min_{\forall k}~\frac{1}{N}\sum_{n=1}^N \alpha_k[n]\left(R_{k,u}[n]-R_{k,e}[n]\right)\label{PF0a}\\
%&\mathrm{s.t.}~\sum_{k=1}^K\sum_{m=1,m\neq k}^Kx_k[n]\xi_{k,m}^*[n]\geq 1-\varepsilon,~\forall n,\label{PF1b}\\
&\mathrm{s.t.}~\sum_{k=1}^K\alpha_k[n]p_{k}^{ro}[n]\leq \epsilon_r,~\forall n,\label{PF0b}\\
&~~~~~\sum_{k=1}^K\alpha_k[n]p_{k}^{so}[n]\leq \epsilon_s,~\forall n,\label{PF0c}\\
&~~~~~\sum_{k=1}^K \alpha_k[n]\leq 1, ~\forall n,\label{PF0d}\\
&~~~~~\alpha_k[n]\in\{0,1\},~\forall k, n,\label{PF0e}\\
&~~~~~P_{u}[n]\leq P_{\max}^u,~\forall n,\label{PF0f}\\
%&~~~~~P_{k}[n]\leq P_{\max}^s,~\forall k,n,\label{PF1e}\\
&~~~~~\mathbf{q}_u[1]=\mathbf{q}_u[N],~\label{PF0g}\\
&~~~~~\|\mathbf{q}_u[n+1]-\mathbf{q}_u[n]\|\leq V_{\max}\delta_t,~n\in\mathcal{N}\setminus\{N\}.\label{PF0h}
\end{align}
\end{subequations}

We note that the term $\frac{1}{N}\sum_{n=1}^N \alpha_k[n]\left(R_{k,u}[n]-R_{k,e}[n]\right)$ in the objective function \eqref{PF0a} denotes the $k$-th SN's ASR (i.e., average secrecy rate) over $N$ time slots,
\eqref{PF0b} is the ROP (i.e., reliability outage probability) constraint, where $\epsilon_r$ is the maximum allowable ROP,
\eqref{PF0c} is the SOP (i.e., secrecy outage probability) constraint, where $\epsilon_s$ is the maximum tolerable SOP determining the required security level.
In addition, \eqref{PF0d} and \eqref{PF0e} are SN scheduling constraints, which ensure that at most one SN is scheduled at each time slot, while \eqref{PF0f} is the UAV's AN transmit power constraint. \eqref{PF0g} and \eqref{PF0h} are the UAV's mobility constraints. We note that the ROP and SOP constraints are to guarantee the reliability and security of the transmission from the scheduled SN to the UAV, respectively, which are the two main constraints in our considered optimization problem. In order to facilitate solving the formulated optimization problem, we first derive the analytic expressions for $p_{k}^{ro}[n]$ and $p_{k}^{so}[n]$ in the following lemma.
\begin{lemma}\label{lemma1}
The analytic expressions for $p_{k}^{ro}[n]$ defined in \eqref{ROP} and $p_{k}^{ro}[n]$ defined in \eqref{SOP} are given by
\begin{align}\label{ROP1}
&p_{k}^{ro}[n]=\exp\left[\frac{-1}{\rho P_u[n]\lambda_{u,u}}\left(\frac{\frac{\beta_0 P^k_s[n]}{\|\mathbf{q}_u[n]-\mathbf{w}_k\|^2+H^2} }{2^{R_{k,u}[n]}-1}-\sigma_u^2\right)\right],
\end{align}
and
\begin{align}\label{SOP1}
&p_{k}^{so}[n]=1\!-\!\prod_{\!m\in \mathcal{K}\setminus \{k\}\!}\left[1\!-\!\exp\left(\frac{\frac{\beta_0 P_u[n]}{\|\mathbf{q}_u[n]-\mathbf{w}_m\|^2+H^2}+\sigma_m^2}{\frac{-P^k_s[n]\lambda_{k,m}}{2^{R_{k,e}[n]}-1}}\right)\right],
\end{align}
respectively.
\end{lemma}
\begin{IEEEproof}
The detailed proof is provided in Appendix \ref{App_0}.
\end{IEEEproof}

\begin{remark}\label{remark1}
We first observe that our objective function detailed in \eqref{PF0a} and $p_k^{ro}[n]$ in \eqref{ROP1} are increasing function of $R_{k,u}[n]$. As such, $\sum_{k=1}^K\alpha_k[n]p_k^{ro}[n]= \epsilon_r$ must be satisfied to maximize the ASR of each SN. We recall that at most one SN is scheduled at each time slot. As such, when SN $k$ is scheduled, $\sum_{k=1}^K\alpha_k[n]p_k^{ro}[n]= \epsilon_r$ is equivalent to $p_k^{ro}[n]= \epsilon_r$. Thus, the transmission rate from the $k$-th SN to the UAV can be written as a function of $\epsilon_r$, given by
\begin{align}\label{R_k}
R_{k,u}[n]=\log_2\left(1+\frac{\frac{\beta_0 P^k_s[n]}{\|\mathbf{q}_u[n]-\mathbf{w}_k\|^2+H^2}}{-\rho P_u[n]\lambda_{u,u}\ln\epsilon_r+\sigma_u^2}\right).
\end{align}
\end{remark}
We also observe that as the quality of the channel $h_{k,u}[n]$ increases, $R_k[n]$ detailed in \eqref{R_k} increases, while as the quality of the channel $h_{u,m}[n]$ increases, the $p_{k}^{so}[n]$ in \eqref{SOP1} decreases.
We note that $R_k[n]$ and $p_{k}^{so}[n]$ decrease with the AN transmit power $P_u[n]$. Thus, the UAV's trajectory and AN transmit power should be carefully designed to balance the transmission rate and the communication security of our considered system.

Following \eqref{R_k}, $(\mathbf{P1})$ can be equivalently reformulated as

\begin{subequations}\label{PF1}
\begin{align}
&(\mathbf{P2}):\max_{\substack{\mathbf{Q},\mathbf{A}\\
\mathbf{P}_{\mathrm{U}},\mathbf{R}_{\mathrm{E}}}}\min_{\forall k}~\frac{1}{N}\sum_{n=1}^N \alpha_k[n]\left(R_{k,u}[n]-R_{k,e}[n]\right)\label{PF1a}\\
%&\mathrm{s.t.}~\sum_{k=1}^K\sum_{m=1,m\neq k}^Kx_k[n]\xi_{k,m}^*[n]\geq 1-\varepsilon,~\forall n,\label{PF1b}\\
&\mathrm{s.t.}~\sum_{k=1}^K\alpha_k[n]p_{k}^{so}[n]\leq \epsilon_s,~\forall n,\label{PF1b}\\
&~~~~~\sum_{k=1}^K \alpha_k[n]\leq 1, ~\forall n,\label{PF1c}\\
&~~~~~\alpha_k[n]\in\{0,1\},~\forall k, n,\label{PF1d}\\
&~~~~~P_{u}[n]\leq P_{\max}^u,~\forall n,\label{PF1e}\\
%&~~~~~P_{k}[n]\leq P_{\max}^s,~\forall k,n,\label{PF1e}\\
&~~~~~\mathbf{q}_u[1]=\mathbf{q}_u[N],~\label{PF1f}\\
&~~~~~\|\mathbf{q}_u[n+1]-\mathbf{q}_u[n]\|\leq V_{\max}\delta_t,~n\in\mathcal{N}\setminus\{N\},\label{PF1g}
\end{align}
\end{subequations}
where $R_{k,u}[n]$ in objective function \eqref{PF1a} is defined in \eqref{R_k}.

%where the term $\frac{1}{N}\sum_{n=1}^N \alpha_k[n]\left(R_{k,u}[n]-R_{k,e}[n]\right)$ in objective function \eqref{PF1a} denotes $k$-th SN's average secrecy rate over the $N$ time slots and $R_{k,u}[n]$ defined in \eqref{RO1} denotes the transmission rate of SN $k$.
%\eqref{PF1b} and \eqref{PF1c} are the ROP constraint and the SOP constraint, respectively, where $\epsilon_r$ is the maximum allowable ROP from SN $k$ to the UAV at the $n$-th time slot and $\epsilon_s$ is the maximum tolerable SOP determining the required security level.
%In addition, \eqref{PF1d} and \eqref{PF1e} are SN scheduling constraints, which ensure that at most one SN is scheduled at each time slot. \eqref{PF1f} is the UAV's AN transmit power constraint, while \eqref{PF1g} and \eqref{PF1h} are the UAV's mobility constraints.

We note that the constraint \eqref{PF1c}, the AN transmit power constraint \eqref{PF1e}, the mobility constraints \eqref{PF1f} and \eqref{PF1g} are convex, while the objective function \eqref{PF1a} and the SOP constraint \eqref{PF1b} are highly non-convex. Furthermore, the SN scheduling variables $\alpha_k[n],\forall k,n$, are binary, and the optimization variables are closely coupled in the objective function \eqref{PF1a} and the SOP constraint \eqref{PF1b}. As such, ($\mathbf{P2}$) is a mixed-integer non-convex optimization problem. We note that finding the global optimal solution to this problem usually requires a high-complexity exhaustive search, which is impractical. In the following section, we develop a novel P-SCA algorithm, which enables us to find a local optimal solution to $(\mathbf{P2})$ within a polynomial time period.

%========================================================================================
\section{UAV's Secure Data collection design}
In this section, we jointly design the UAV's trajectory, AN transmit power and the redundancy rate as well as the SN scheduling with the aim to solve the formulated optimization problem $(\mathbf{P2})$. We first transform $(\mathbf{P2})$ into a convex optimization problem, then we further convert it into a SOCP, and finally we develop a P-SCA algorithm to solve it.
\newcounter{mytempeqncnt1}
\begin{figure*}[tp]
\normalsize
\setcounter{mytempeqncnt1}{\value{equation}}
\setcounter{equation}{26}
\begin{align}\label{PS10}
&\log_2\left(1+\frac{\frac{\beta_0 P^k_s[n]}{\|\mathbf{\tilde{q}}_u[n]-\mathbf{w}_k\|^2+H^2}}{-\rho \tilde{P}_u[n]\lambda_{u,u}\ln\epsilon_r+\sigma_u^2}\right)+
\frac{\frac{-\beta_0P^k_s[n]\left(\|\mathbf{q}_u[n]-\mathbf{w}_k\|^2-\|\mathbf{\tilde{q}}_u[n]-\mathbf{w}_k\|^2\right)}{\|\mathbf{\tilde{q}}_u[n]-\mathbf{w}_k\|^2+H^2}}{\left(\left(\|\mathbf{\tilde{q}}_u[n]-\mathbf{w}_k\|^2+H^2\right)\left(-\rho \tilde{P}_u[n]\lambda_{u,u}\ln\epsilon_r+\sigma_u^2\right)+\beta_0 P^k_s[n]\right)\ln2}\nonumber\\
&~~~~~~~~~~~~~~+\frac{\frac{-\beta_0P^k_s[n]\left(-\rho P_u[n]\lambda_{u,u}\ln\epsilon_r+\rho \tilde{P}_u[n]\lambda_{u,u}\ln\epsilon_r\right)}{-\rho \tilde{P}_u[n]\lambda_{u,u}\ln\epsilon_r+\sigma_u^2}}{\left(\left(\|\mathbf{\tilde{q}}_u[n]-\mathbf{w}_k\|^2+H^2\right)\left(-\rho \tilde{P}_u[n]\lambda_{u,u}\ln\epsilon_r+\sigma_u^2\right)+\beta_0 P^k_s[n]\right)\ln2}-R_{k,e}[n]\geq\mu_k[n],~\forall k,n.
\end{align}
\setcounter{equation}{\value{mytempeqncnt1}}
\hrulefill
\vspace*{4pt}
\end{figure*}
\subsection{Transform $(\mathbf{P2})$ into a Convex Optimization Problem}

In this subsection, we aim to transform the mixed-integer non-convex problem $(\mathbf{P2})$ into a convex optimization problem. Specifically, we first introduce a penalty factor to transform $(\mathbf{P2})$ into a continuous optimization problem. Then, we employ the first-order restrictive method to convert the continuous optimization problem into a convex optimization problem.

To proceed, we note that \eqref{PF1d} can actually be equivalently rewritten as the following continuous constraint, i.e.,
\begin{subequations}\label{PS1}
\begin{align}
\alpha_k[n]-\alpha_k[n]^2\leq 0,~\forall k, n,\label{PS1a}\\
0\leq \alpha_k[n]\leq 1,~\forall k, n,\label{PS1b}
\end{align}
\end{subequations}
where $\alpha_k[n]\leq0$ or $\alpha_k[n]\geq 1$ must hold in \eqref{PS1a}. Combining  \eqref{PS1a} and \eqref{PS1b}, we have $\alpha_k[n]=0$ or $\alpha_k[n]=1$. We note that \eqref{PS1} can be further simplified as
\begin{subequations}\label{PS2}
\begin{align}
\sum_{n=1}^N\sum_{k=1}^K\left(\alpha_k[n]-\alpha_k[n]^2\right)\leq 0,\label{PS2a}\\
0\leq \alpha_k[n]\leq 1,~\forall k, n.\label{PS2b}
\end{align}
\end{subequations}
We note that the number of constraints in \eqref{PS2} is fewer than that in \eqref{PS1}, which can significantly reduce the computational complexity of solving $(\mathbf{P2})$.
Introducing a slack variable $\eta$ and replacing the binary constraint \eqref{PF1d} by \eqref{PS2}, problem (P2) can be equivalently rewritten as
\begin{subequations}\label{PF2_1}
\begin{align}
&(\mathbf{P2.1}):\max_{\substack{\mathbf{Q},\mathbf{A}, \eta\\
\mathbf{P}_{\mathrm{U}},\mathbf{R}_{\mathrm{E}}}}~\eta\\
&\mathrm{s.t.}~\frac{1}{N}\sum_{n=1}^N \alpha_k[n]\left(R_{k,u}[n]-R_{k,e}[n]\right)\geq \eta,~\forall k,\label{PF2_1b}\\
&~~~~~\eqref{PF1b},\eqref{PF1c},\eqref{PF1e},\eqref{PF1f},\eqref{PF1g},\eqref{PS2a},\eqref{PS2b}.
\end{align}
\end{subequations}
Although we have transformed the original mixed-integer optimization problem $(\mathbf{P2})$ into the continuous optimization problem $(\mathbf{P2.1})$, it is still non-convex due to the constraints \eqref{PF1b} and \eqref{PS2a} together with \eqref{PF2_1b}. In general, we can apply the first-order restrictive approximation to transform the problem $(\mathbf{P2.1})$ into a convex optimization problem and then employ the SCA technique to solve the resultant problem.
However, direct applying the SCA technique will make it difficult to find the initial feasible solution due to the joint existence of \eqref{PS2a} and \eqref{PS2b}. To overcome this issue, we introduce a slack variable $\phi$ to extend feasible set of constraint \eqref{PS2} and develop a penalty method to add this slack variable $\phi$ into the objective function. Then, the resultant optimization problem is given by
\begin{subequations}\label{PF2_2}
\begin{align}
&(\mathbf{P2.2}):\max_{\substack{\mathbf{Q},\mathbf{A}, \eta\\
\mathbf{P}_{\mathrm{U}},\mathbf{R}_{\mathrm{E}},\phi}}~\eta-\omega\phi\\
&\mathrm{s.t.}~\sum_{n=1}^N\sum_{k=1}^K\left(\alpha_k[n]-\alpha_k[n]^2\right)\leq \phi,\label{PF2_2b}\\
&~~~~~\eqref{PF1b},\eqref{PF1c},\eqref{PF1e},\eqref{PF1f},\eqref{PF1g},\eqref{PS2b},\eqref{PF2_1b},
\end{align}
\end{subequations}
where $\omega>0$ is a penalty parameter. We note that $\phi$ is to be minimized in $(\mathbf{P2.2})$ and $\phi=0$ immediately implies that
$(\mathbf{P2.2})$ is equivalent to $(\mathbf{P2.1})$. In the following, we first apply the first-order restrictive approximation to transform the non-convex SN scheduling constraint \eqref{PF2_2b}, constraint \eqref{PF2_1b}, and SOP constraint \eqref{PF1b} into convex constraints.
Then, we develop a P-SCA algorithm to solve the achieved optimization problem.

\subsubsection{The constraint \eqref{PF2_2b}}

We observe that each summation term on the left hand side (LHS) of \eqref{PF2_2b} is in the form of a linear function minus a quadratic convex function. This special form allows us to apply the first-order restrictive approximation to transform the constraint \eqref{PF2_2b} into a convex constraint. We note that any convex function is lower bounded by its first-order approximation. Thus, for given feasible points $\tilde{\alpha}_k[n],\forall k,n$, \eqref{PF2_2b} can be rewritten as
\begin{align}\label{PS3}
\sum_{n=1}^N\sum_{k=1}^K\left(\alpha_k[n]+\tilde{\alpha}_k[n]^2-2\tilde{\alpha}_k[n]\alpha_k[n]\right)\leq \phi.
\end{align}
We note that \eqref{PS3} is a convex constraint due to the fact that it is a linear function with respect to the SN scheduling variable $\alpha_k[n]$ and slack variable $\phi$.
We also note that the constraint \eqref{PS3} is stricter than the constraint \eqref{PF2_2b}. As such, any solution satisfying \eqref{PS3} can also guarantee \eqref{PF2_2b}.

\subsubsection{The constraint \eqref{PF2_1b}}

Substituting $R_{k,u}[n]$ defined in \eqref{R_k} into \eqref{PF2_1b}, the resultant constraint is given by
\begin{align}\label{PS4}
&\sum_{n=1}^N \alpha_k[n]\left[\log_2\left(1+\frac{\frac{\beta_0 P^k_s[n]}{\|\mathbf{q}_u[n]-\mathbf{w}_k\|^2+H^2}}{-\rho P_u[n]\lambda_{u,u}\ln\epsilon_r\!+\!\sigma_u^2}\right)\!-\!R_{k,e}[n]\right]\nonumber\\
&~~~~~~~~~~~~~~~~~~~~~~~~~~~~~~~~~~~~~~~~~~~~~~~~~\geq N\eta,~\forall k.
\end{align}
We observe that the constraint \eqref{PS4} is a non-convex constraint due to the fact that the UAV trajectory $\mathbf{q}_u[n]$ and AN transmit power $P_u[n]$ together with SN scheduling $\alpha_k[n]$ in constraint \eqref{PS4} are closely coupled, which makes it difficult to handle directly. To facilitate tackling this constraint, we first introduce slack variables $\mu_k[n]$, $\forall k,n$, and then we rewrite \eqref{PS4} as
\begin{subequations}\label{PS5}
\begin{align}
&\sum_{n=1}^N \alpha_k[n]\mu_k[n]\geq N\eta, ~\forall k,\label{PS5a}\\
&\log_2\left(1+\frac{\frac{\beta_0 P^k_s[n]}{\|\mathbf{q}_u[n]-\mathbf{w}_k\|^2+H^2}}{-\rho P_u[n]\lambda_{u,u}\ln\epsilon_r+\sigma_u^2}\right)\!-\!R_{k,e}[n]\geq \mu_k[n],\nonumber\\
&~~~~~~~~~~~~~~~~~~~~~~~~~~~~~~~~~~~~~~~~~~~~~~~~~~~~~~~\forall k,n.\label{PS5b}
\end{align}
\end{subequations}
We note that, although \eqref{PS5a} and \eqref{PS5b} are still non-convex, they are easier to be tackled than the original constraint \eqref{PF2_1b}. In the following, we handle the constraints \eqref{PS5a} and \eqref{PS5b} based on their special structures. To proceed, we first observe that \eqref{PS5a} can be equivalently rewritten as
\begin{align}\label{PS6}
\sum_{n=1}^N \left[(\alpha_k[n]+\mu_k[n])^2\!-\!(\alpha_k[n]-\mu_k[n])^2\right]\geq 4N\eta,\forall k.
\end{align}
We note that the LHS of \eqref{PS6} is in the form of the difference of two quadratic convex functions. As such, for given feasible points $\tilde{\alpha}_k[n]$ and $\tilde{\mu}_k[n]$, the first-order restrictive approximation of \eqref{PS6} is given by
\begin{align}\label{PS7}
&\sum_{n=1}^N \big[2(\tilde{\alpha}_k[n]+\tilde{\mu}_k[n])(\alpha_k[n]+\mu_k[n])\nonumber\\ &~-(\tilde{\alpha}_k[n]+\tilde{\mu}_k[n])^2-(\alpha_k[n]-\mu_k[n])^2\big]\geq 4N\eta,\forall k.
\end{align}
We note that the constraint \eqref{PS7} is a convex constraint and it is stricter than \eqref{PS6}. Furthermore, \eqref{PS6} and \eqref{PS7} are equivalent at the given feasible points $\tilde{\alpha}_k[n]$ and $\tilde{\mu}_k[n]$.

To facilitate dealing with the non-convex constraint \eqref{PS5b}, we first define a function $f_1(x_1,x_2)$, which is given by
\begin{align}\label{PS8}
f_1(x_1,x_2)=\log_2\left(1+\frac{c}{x_1x_2}\right),
\end{align}
where $c\geq 0$, $x_1>0$ and $x_2>0$. We note that $f_1(x_1,x_2)$ is jointly convex with respect to $x_1$ and $x_2$\cite{zhou2019twc}. Following the fact that any convex function is lower bounded by its first order approximation, we have the following inequality
\begin{align}\label{PS9}
&f_1(x_1,x_2)\geq\log_2\left(1+\frac{c}{\tilde{x}_1\tilde{x}_2}\right)\nonumber\\
&~~~~~~~~~~+\frac{-c(x_1-\tilde{x}_1)}{\tilde{x}_1(\tilde{x}_1\tilde{x}_2+c)\ln2}+\frac{-c(x_2-\tilde{x}_2)}{\tilde{x}_2(\tilde{x}_1\tilde{x}_2+c)\ln2},
\end{align}
where $\tilde{x}_1$ and $\tilde{x}_2$ are first-order Taylor expansion points.

Now, we return to the constraint \eqref{PS5b}. We observe that the term $\log_2\Big(1+\frac{\frac{\beta_0 P^k_s[n]}{\|\mathbf{q}_u[n]-\mathbf{w}_k\|^2+H^2}}{-\rho P_u[n]\lambda_{u,u}\ln\epsilon_r+\sigma_u^2}\Big)$ in \eqref{PS5b} is in a similar form as
\eqref{PS8}. Specifically, $\beta_0P^k_s[n]\geq 0$, $\|\mathbf{q}_u[n]-\mathbf{w}_k\|^2+H^2>0$, and $-\rho P_u[n]\lambda_{u,u}\ln\epsilon_r+\sigma_u^2>0$ must hold.
Following the convexity of $f_1(x_1,x_2)$,
the term $\log_2\Big(1+\frac{\frac{\beta_0 P^k_s[n]}{\|\mathbf{q}_u[n]-\mathbf{w}_k\|^2+H^2}}{-\rho P_u[n]\lambda_{u,u}\ln\epsilon_r+\sigma_u^2}\Big)$ is jointly convex with respect to $\|\mathbf{q}_u[n]-\mathbf{w}_k\|^2+H^2$ and $-\rho P_u[n]\lambda_{u,u}\ln\epsilon_r+\sigma_u^2$.
 We replace $c$, $x_1$, $x_2$, $\tilde{x}_1$, and $\tilde{x}_2$ in \eqref{PS9} with $\beta_0P^k_s[n]$, $\|\mathbf{q}_u[n]-\mathbf{w}_k\|^2+H^2$, $-\rho P_u[n]\lambda_{u,u}\ln\epsilon_r+\sigma_u^2$, $\|\mathbf{\tilde{q}}_u[n]-\mathbf{w}_k\|^2+H^2$, and $-\rho \tilde{P}_u[n]\lambda_{u,u}\ln\epsilon_r+\sigma_u^2$, respectively. Then, the first-order restrictive approximation of the constraint \eqref{PS5b} is given by \eqref{PS10}, which is presented at the top of previous page.
We note that \eqref{PS10} is a convex constraint.

\setcounter{equation}{27}

So far, we have transformed the non-convex constraint \eqref{PF2_1b} into the convex constraints \eqref{PS7} and \eqref{PS10}.

\subsubsection{The SOP constraint \eqref{PF1b}}

We first substitute \eqref{SOP1} into the SOP constraint \eqref{PF1b} and we have
\begin{align}\label{PS11}
&\sum_{k=1}^K\alpha_k[n]\left[1\!-\!\!\!\!\prod_{\!m\in \mathcal{K}\setminus \!\{k\}\!\!}\left[1\!-\!\exp\!\left(\frac{\frac{\beta_0 P_u[n]}{\|\mathbf{q}_u[n]-\mathbf{w}_m\|^2+H^2}\!+\!\sigma_m^2}{\frac{-P^k_s[n]\lambda_{k,m}}{2^{R_{k,e}[n]}-1}}\!\right)\right]\right]\nonumber\\
&~~~~~~~~~~~~~~~~~~~~~~~~~~~~~~~~~~~~~~~~~~~~~~~~~~~~\leq \epsilon_s,\forall n.
\end{align}
We note that the main challenge to tackle the constraint \eqref{PS11} arises from the fact that the expression of \eqref{PS11} is of a high complexity and the optimization variables in constraint \eqref{PS11} are closely coupled. To overcome this challenge, we introduce slack variables $\nu_k[n]$, $\forall k,n$, and rewrite the constraint \eqref{PS11} as
\begin{subequations}\label{PS12}
\begin{align}
&\sum_{k=1}^K \alpha_k[n]\nu_k[n]\leq \epsilon_s, ~\forall n,\label{PS12a}\\
&1\!-\!\prod_{m\in \mathcal{K}\setminus \{k\}}\left[1-\exp\left(\frac{\frac{\beta_0 P_u[n]}{\|\mathbf{q}_u[n]-\mathbf{w}_m\|^2+H^2}+\sigma_m^2}{\frac{-P^k_s[n]\lambda_{k,m}}{2^{R_{k,e}[n]}-1}}\right)\right]\leq \nu_k[n],\nonumber\\
&~~~~~~~~~~~~~~~~~~~~~~~~~~~~~~~~~~~~~~~~~~~~~~~~~~~~~~~~\forall k,n.\label{PS12b}
\end{align}
\end{subequations}
We note that \eqref{PS12a} and \eqref{PS12b} are still non-convex constraints. However, they are relatively easier to tackle compared with the original SOP constraint \eqref{PS11}.
In the following, we focus on handling the non-convex constraints \eqref{PS12a} and \eqref{PS12b}.

We first note that \eqref{PS12a} can be equivalently rewritten as
\begin{align}\label{PS13}
\sum_{k=1}^K \left[(\alpha_k[n]+\nu_k[n])^2\!-\!(\alpha_k[n]-\nu_k[n])^2\right]\leq 4\epsilon_s,\forall n.
\end{align}
Similar to \eqref{PS6}, for given feasible points $\tilde{\alpha}_k[n]$ and $\tilde{\nu}_k[n]$, $\forall k,n$, the first-order restrictive approximation of \eqref{PS13} is given by
\begin{align}\label{PS14}
&\sum_{k=1}^K \big[-2(\tilde{\alpha}_k[n]-\tilde{\nu}_k[n])(\alpha_k[n]-\tilde{\alpha}_k[n]+\tilde{\nu}_k[n]-\nu_k[n])\nonumber\\
&+(\alpha_k[n]+\nu_k[n])^2\!-\!(\tilde{\alpha}_k[n]-\tilde{\nu}_k[n])^2\big]\leq 4\epsilon_s,\forall n.
\end{align}

For the non-convex constraint \eqref{PS12b}, we first observe that \eqref{PS12b} is in the form of the product of multiple exponential functions, which is generally difficult to handle. We recall that
the function $\prod_{k=1}^K\frac{1}{e_k}$ is a convex function for $e_k>0$\cite{Boyd}. Following this fact, we introduce slack variables $\theta_{k,m}[n]$, $\forall k,n,m\in \mathcal{K}\setminus \{k\}$, and rewrite \eqref{PS12b} as
\begin{subequations}\label{PS15}
\begin{align}
&\prod_{m=1,m\neq k}^K\frac{1}{\theta_{k,m}[n]}\geq 1-\nu_k[n],\forall k,n.\label{PS15a}\\
&1-\exp\left(\frac{\frac{\beta_0 P_u[n]}{\|\mathbf{q}_u[n]-\mathbf{w}_m\|^2+H^2}+\sigma_m^2}{\frac{-P^k_s[n]\lambda_{k,m}}{2^{R_{k,e}[n]}-1}}\right)\geq \frac{1}{\theta_{k,m}[n]}, \nonumber\\
&~~~~~~~~~~~~~~~~~~~~~~~~~~~~~~~~~~~~~~\forall k,n,m\in \mathcal{K}\setminus \{k\}.\label{PS15b}
\end{align}
\end{subequations}
We note that \eqref{PS15a} is in the form of the super-level set of a convex function, which is non-convex. In addition, \eqref{PS15b} is also a non-convex constraint, which is difficult to tackle directly due to the fact that the expression of \eqref{PS15b} is of a high complexity. In the following, we aim to transform \eqref{PS15a} and \eqref{PS15b} into convex constraints.

To handle the non-convex constraint \eqref{PS15a}, we first present the following inequality by performing the first-order Taylor approximation at points $\tilde{\theta}_{k,m}[n]$, $m\in \mathcal{K}\setminus \{k\}$, i.e.,
\begin{align}\label{PS15_1}
\prod_{m=1,m\neq k}^K\frac{1}{\theta_{k,m}[n]}&\geq f_2\left(\theta_{k,m}[n],\tilde{\theta}_{k,m}[n]\right)\nonumber\\
&\triangleq\frac{\sum\limits_{m=1,m\neq k}^K\frac{-\theta_{k,m}[n]}{\tilde{\theta}_{k,m}[n]}+K}{\prod\limits_{m=1,m\neq k}^K\frac{1}{\tilde{\theta}_{k,m}[n]}},\forall k,n.
\end{align}
We note that the inequality \eqref{PS15_1} is due to the fact that the convexity of the function $\prod_{m=1,m\neq k}^K\frac{1}{\theta_{k,m}[n]}$. Following \eqref{PS15_1}, the first-order restrictive approximation of the constraint \eqref{PS15a} is given by
\begin{align}\label{PS15_2}
f_2\left(\theta_{k,m}[n],\tilde{\theta}_{k,m}[n]\right)\geq 1-\nu_k[n],\forall k,n.
\end{align}
We note that the constraint \eqref{PS15_2} is linear with respect to the introduced slack variables $\theta_{k,m}[n]$ and $\nu_k[n]$ for given feasible point $\tilde{\theta}_{k,m}[n]$. As such, the constraint \eqref{PS15_2} is convex.

Now, we turn our attention to the non-convex constraint \eqref{PS15b}. To facilitate processing the constraint \eqref{PS15b}, we first introduce slack variables $\varsigma_m[n]$ and $\tau_k[n]$, $\forall k,n,m\in \mathcal{K}\setminus \{k\}$, and then rewrite \eqref{PS15b} as
\begin{subequations}\label{PS16}
\begin{align}
&\exp\left(\frac{-\sigma^2}{P^k_s[n]\lambda_{k,m}}\sqrt{\varsigma_m[n]\tau_k[n]}\right)\nonumber\\
&~~~~~~~~~~~~~~~~~~\leq1-\frac{1}{\theta_{k,m}[n]},\forall k,n,m\in \mathcal{K}\setminus \{k\},\label{PS16a}\\
&\frac{\frac{\beta_0}{\sigma^2}P_u[n]}{\|\mathbf{q}_u[n]\!-\!\mathbf{w}_m\|^2\!+\!H^2}\!+\!1\geq \sqrt{\varsigma_m[n]},\forall n,m\in \mathcal{K}\!\setminus \{k\}\!,\label{PS16b}\\
&2^{R_{k,e}[n]}-\sqrt{\tau_k[n]}\geq 1,\forall k,n.\label{PS16c}
\end{align}
\end{subequations}
%We observe that the constraint \eqref{PS16a} is convex, while \eqref{PS16b} and \eqref{PS16c} are still non-convex. However, we can applying the first-order restrictive approximation to transform them into convex constraints. In the following, we present the detailed transformation processes.

We note that the constraint \eqref{PS16a} is convex due to the following three facts.
Firstly, $\sqrt{\varsigma_m[n]\tau_k[n]}$ is a geometric mean function, which is a joint concave function with respect to the introduced slack variables $\varsigma_m[n]$ and $\tau_k[n]$. Secondly, if a function $g(x)$ is convex then $\exp{(g(x))}$ is convex, i.e., the exponential function satisfies convexity-preserving operation. Thirdly, $1-\frac{1}{\theta_{k,m}[n]}$ in the right hand side (RHS) of \eqref{PS16a} is a concave function.

%We observe that the RHS of the \eqref{PS16a} is a concave function for $K\geq 2$. We note that any concave function is upper bounded by its first-order approximation. As such, for given feasible point $\tilde{\theta}_{k,m}[n]$, $\forall k,n,m\in \mathcal{K}\setminus \{k\}$, the restrictive approximation of the constraint \eqref{PS16a} is given by
%\begin{align}\label{PS17}
%&1-\exp\left(\frac{-\sigma^2}{P^k_s[n]\lambda_{k,m}}\left(\sqrt{\varsigma_m[n]\tau_k[n]}+\sqrt{\tau_k[n]}\right)\right)\nonumber\\
%&\geq(\tilde{\theta}_{k,m}[n])^{\frac{1}{K-1}}+\frac{1}{K-1}(\tilde{\theta}_{k,m}[n])^{\frac{2-K}{K-1}}(\theta_{k,m}[n]-\tilde{\theta}_{k,m}[n])\nonumber\\
%&~~~~~~~~~~~~~~~~~~~~~~~~~~~~~~~~~~~~~~~~\forall k,n,m\in \mathcal{K}\setminus \{k\}.
%\end{align}
%We note that the RHS of the constraint \eqref{PS17} is a linear function with respect to the slack variable $\theta_{k,m}[n]$ for given feasible point $\tilde{\theta}_{k,m}[n]$. As such, the constraint \eqref{PS17} is a convex constraint.
%Furthermore, the constraint \eqref{PS17} is stricter than the original constraint \eqref{PS16a}, which implies that any feasible solution to the constraint \eqref{PS17} is also feasible to \eqref{PS16a}.

We observe that the LHS of the constraint \eqref{PS16b} is not a convex function with respect to the optimization variables and the RHS of \eqref{PS16b} is a concave function. Thus, \eqref{PS16b} is non-convex. In addition, the constraint \eqref{PS16c} is non-convex due to the fact that the super-level of a convex function is non-convex.
In the following, we present the detailed processes of tackling \eqref{PS16b} and \eqref{PS16c}.

To facilitate processing the constraint \eqref{PS16b}, we first define a function $f_3(x_1,x_2)$, which is given by
\begin{align}\label{PS18}
f_3(x_1,x_2)=\frac{c}{x_1x_2},
\end{align}
where $c\geq0$, $x_1>0$ and $x_2>0$. We can see that $f_3(x_1,x_2)$ is jointly convex with respect to $x_1$ and $x_2$\cite{zhou2019twc}. As a result, we have the following inequality
\begin{align}\label{PS19}
&f_3(x_1,x_2)\geq \frac{c}{\tilde{x}_1\tilde{x}_2}-\frac{c(x_1-\tilde{x}_1)}{\tilde{x}_1^2\tilde{x}_2}-\frac{c(x_2-\tilde{x}_2)}{\tilde{x}_1\tilde{x}_2^2},
\end{align}
where $\tilde{x}_1$ and $\tilde{x}_2$ are first-order Taylor expansion points. In order to apply the results of \eqref{PS19}, we rearrange \eqref{PS16b} as
%\begin{align}\label{PS20}
%&\frac{\|\mathbf{q}_u[n]-\mathbf{w}_m\|^2+H^2}{\frac{\beta_0}{\sigma^2}P_u[n]}\leq \sqrt{\frac{1}{\varsigma_m[n]}},\forall n,m\in \mathcal{K}\setminus \{k\},
%\end{align}
\begin{align}\label{PS20}
&\frac{\frac{\beta_0}{\sigma^2}}{\frac{1}{P_u[n]}\left(\|\mathbf{q}_u[n]-\mathbf{w}_m\|^2+H^2\right)}+1-\sqrt{\varsigma_m[n]}\nonumber\\
&~~~~~~~~~~~~~~~~~~~~~~~~~~~~~~~~~~\geq 0,\forall n,m\in \mathcal{K}\setminus \{k\}.
\end{align}
We note that the term $\frac{\frac{\beta_0}{\sigma^2}}{\frac{1}{P_u[n]}\left(\|\mathbf{q}_u[n]-\mathbf{w}_m\|^2+H^2\right)}$ in \eqref{PS20} is in a similar form as the function $f_3(x_1,x_2)$. Consequently, we replace $c$, $x_1$, $x_2$, $\tilde{x}_1$, and $\tilde{x}_2$ in \eqref{PS19} with $\frac{\beta_0}{\sigma^2}$, $\frac{1}{P_u[n]}$, $\|\mathbf{q}_u[n]-\mathbf{w}_m\|^2+H^2$, $\frac{1}{\tilde{P}_u[n]}$, and $\|\mathbf{\tilde{q}}_u[n]-\mathbf{w}_m\|^2+H^2$, respectively.
Furthermore, the term $-\sqrt{\varsigma_m[n]}$ in \eqref{PS20} is a convex function with respect to the slack variable $\varsigma_m[n]$. We note that any super-level set of a convex function is non-convex. As such, we can linearize the convex function $-\sqrt{\varsigma_m[n]}$ by employing the first-order approximation.
Following the above discussion, \eqref{PS20} can be rewritten as
\begin{align}\label{PS21}
&\frac{\frac{\beta_0}{\sigma^2}\tilde{P}_u[n]}{\|\mathbf{\tilde{q}}_u[n]-\mathbf{w}_m\|^2+H^2}-\frac{\frac{\beta_0}{\sigma^2}(\frac{1}{P_u[n]}-\frac{1}{\tilde{P}_u[n]})}{\frac{1}{(\tilde{P}_u[n])^2}(\|\mathbf{\tilde{q}}_u[n]-\mathbf{w}_m\|^2+H^2)}\nonumber\\
&-\frac{\frac{\beta_0}{\sigma^2}(\|\mathbf{q}_u[n]-\mathbf{w}_m\|^2-\|\mathbf{\tilde{q}}_u[n]-\mathbf{w}_m\|^2)}{\frac{1}{\tilde{P}_u[n]}(\|\mathbf{\tilde{q}}_u[n]-\mathbf{w}_m\|^2+H^2)^2}
+1-\sqrt{\tilde{\varsigma}_m[n]}\nonumber\\
&-\frac{1}{2}(\tilde{\varsigma}_m[n])^{\frac{-1}{2}}(\varsigma_m[n]-\tilde{\varsigma}_m[n])\geq 0,\forall n,m\in \mathcal{K}\setminus \{k\},
\end{align}
where $\tilde{P}_u[n]$, $\tilde{q}_u[n]$, and $\tilde{\varsigma}_m[n]$ are given feasible points.

Finally, we focus on tackling the constraint \eqref{PS16c}. For given feasible points $\tilde{R}_{k,e}[n]$ and $\tilde{\tau}_k[n]$, the first-order restrictive approximation of \eqref{PS16b} is given by
\begin{align}\label{PS22}
&2^{\tilde{R}_{k,e}[n]}+2^{\tilde{R}_{k,e}[n]}\ln{2}(R_{k,e}[n]-\tilde{R}_{k,e}[n])-\sqrt{\tilde{\tau}_k[n]}\nonumber\\
&~~~~~~~~~~~~~-\frac{1}{2}(\tilde{\tau}_k[n])^{\frac{-1}{2}}(\tau_k[n]-\tilde{\tau}_k[n])
\geq 1,\forall k,n.
\end{align}

So far, we have transformed the non-convex constraint \eqref{PF1b} into the convex constraints \eqref{PS14}, \eqref{PS15_2}, \eqref{PS16a}, \eqref{PS21} and \eqref{PS22}.

Following the above transformations, we rewrite $(\mathbf{P2.2})$ as
\begin{align}\label{PF2_3}
&(\mathbf{P2.3}):\max_{\substack{\mathbf{Q},\mathbf{A}, \eta,\mathbf{P}_{\mathrm{U}},\mathbf{R}_{\mathrm{E}},\phi\\
\mathbf{U},\mathbf{V},\mathbf{\Theta},\mathbf{S}_1,\mathbf{S}_2}}~\eta-\omega\phi\nonumber\\
&\mathrm{s.t.}~\eqref{PF1c},\eqref{PF1e},\eqref{PF1f},\eqref{PF1g},\eqref{PS2b},\nonumber\\
&~~~~~\eqref{PS3},\eqref{PS7},\eqref{PS10},\eqref{PS14},\eqref{PS15_2},\eqref{PS16a},\eqref{PS21},\eqref{PS22},
\end{align}
where $\mathbf{U}\triangleq\{\mathbf{\mu}_k[n],\forall k,n\}$, $\mathbf{V}\triangleq\{\mathbf{\nu}_k[n],\forall k,n\}$, $\mathbf{\Theta}\triangleq\{\theta_{k,m}[n]$, $\forall k,n,m\in \mathcal{K}\setminus \{k\}\}$, $\mathbf{S}_1\triangleq\{\mathbf{\varsigma}_k[n],\forall k,n\}$, and $\mathbf{S}_2\triangleq\{\mathbf{\tau}_k[n],\forall k,n\}$.
We note that problem $(\mathbf{P2.3})$ is a convex optimization problem, which can be solved by the convex optimization tool, such as CVX\cite{Boyd}.

\subsection{Second-Order Cone Representation}

In the previous subsection, we have transformed the original optimization problem ($\mathbf{P2}$) into the convex optimization problem ($\mathbf{P2.3}$). We observe that problem ($\mathbf{P2.3}$) is categorized as a GNCP (i.e.,  generalized nonlinear convex programming) due to the exponential function involved in the constraint \eqref{PS16a}\cite{Tervo2015Optimal}. We should clarify that problem ($\mathbf{P2.3}$) can indeed be solved using the interior-point solver. However, solving it directly often results in a high computational complexity compared to other standard convex programs such as the SOCP. This motivates us to further transform ($\mathbf{P2.3}$) into a standard convex program. Following this consideration, we observe from problem ($\mathbf{P2.3}$) that the objective function and the constraints are linear or SOC presentable, except the exponential cone constraint \eqref{PS16a}.
As such, in the following we dedicate to converting ($\mathbf{P2.3}$) into a SOCP. To this end, we first present a proposition to reformulate \eqref{PS7}, \eqref{PS10}, \eqref{PS14} and \eqref{PS21} into SOC constraints. Then, we focus on representing the exponential cone constraint \eqref{PS16a} into a SOC form.
\begin{proposition}\label{proposition1}
The constraints \eqref{PS7}, \eqref{PS10} and \eqref{PS14} can be represented as the SOC constraints given by
\begin{align}
&\left\|\alpha_k[1]-\mu_k[1],\cdots, \alpha_k[N]-\mu_k[N], \frac{A_k[n]-1}{2}\right\|
\nonumber\\
&~~~~~~~~~~~~~~~~~~~~~~~~~~~~~~~~~~~~\leq \frac{A_k[n]+1}{2},\forall k,\label{SOC1}\\
&\left\|\sqrt{\frac{\beta_0 P^k_s[n]}{\|\mathbf{\tilde{q}}_u[n]-\mathbf{w}_k\|^2+H^2}}(\mathbf{q}_u[n]-\mathbf{w}_k)^T,\frac{B_k[n]-1}{2}\right\|\nonumber\\
&~~~~~~~~~~~~~~~~~~~~~~~~~~~~~~~~~~~~~~\leq \frac{B_k[n]+1}{2},\forall k,n,\label{SOC2}\\
&\bigg\|\alpha_1[n]+\nu_1[n],\cdots,\alpha_K[n]\!+\!\nu_K[n],\tilde{\alpha}_1[n]-\tilde{\nu}_1[n],\cdots,\nonumber\\
&~~~~~~~~\tilde{\alpha}_K[n]-\tilde{\nu}_K[n],\frac{C_k[n]-1}{2}\bigg\|\leq \frac{C_k[n]+1}{2},\forall n,\label{SOC3}
\end{align}
respectively, where $A_k[n]$, $B_k[n]$ and $C_k[n]$ are defined in \eqref{App_A2}, \eqref{App_A4} and \eqref{App_A6}, respectively.
In addition, the constraint \eqref{PS21} can be reformulated as the following SOC constraints:
 \begin{align}
\left\|(\mathbf{q}_u[n]-\mathbf{w}_m)^T,\frac{D_m[n]-1}{2}\right\|^2&\leq \frac{D_m[n]+1}{2},\label{SOC4}\\
\left\|\frac{P_u[n]-\zeta[n]}{2},1\right\|^2&\leq \frac{P_u[n]+\zeta[n]}{2},\label{SOC5}
\end{align}
where $D_m[n]$ is defined in \eqref{App_A8} and $\zeta[n]$ is an introduced slack variable.
\end{proposition}
\begin{IEEEproof}
The detailed proof is provided in Appendix \ref{App_A}.
\end{IEEEproof}

Now, we turn our attention to the exponential cone constraint \eqref{PS16a}. In the existing work \cite{Tervo2015Optimal}, the exponential function is first expanded using the Taylor series, and then the resultant terms are converted into several SOC constraints. Although a factor is introduced in \cite{Tervo2015Optimal} to control the approximation accuracy, it cannot guarantee that the Taylor series expansion is an upper bound of the original exponential function. This may lead to a solution to the SOCP problem that is not feasible to the original problem.
As such, it is important to develop a restrictive approximation method to express the exponential cone constraint into the SOC form.
In this work, we present a new method to overcome this problem.
To this end, we first present the following inequality\cite{Nguyen2019Energy}
\begin{align}\label{PS34}
\log(x)\geq \log(\tilde{x})+2-\frac{2\sqrt{\tilde{x}}}{\sqrt{x}},
\end{align}
which holds for all $x>0$ and $\tilde{x}>0$. We note that \eqref{PS34} provides a lower bound of the logarithmic function and the equality must hold for $x=\tilde{x}$. In the following, we use this inequality to deal with the exponential cone constraint \eqref{PS16a}.

To proceed, we first equivalently rewrite \eqref{PS16a} as
\begin{align}\label{PS35}
&\frac{-\sigma^2}{P^k_s[n]\lambda_{k,m}}\sqrt{\varsigma_m[n]\tau_k[n]}\leq\log\left(1-\frac{1}{\theta_{k,m}[n]}\right),
\end{align}
$\forall k,n,m\in \mathcal{K}\setminus \{k\}$. Following the inequality \eqref{PS34}, the restrictive approximation for \eqref{PS35} is given by
\begin{align}\label{PS36}
&\frac{\sigma^2}{P^k_s[n]\lambda_{k,m}}\sqrt{\varsigma_m[n]\tau_k[n]}\geq-\log\left(1-\frac{1}{\tilde{\theta}_{k,m}[n]}\right)\nonumber\\
&~~~~~~~~~~~~-2+\frac{2\sqrt{1-\frac{1}{\tilde{\theta}_{k,m}[n]}}}{\sqrt{1-\frac{1}{\theta_{k,m}[n]}}},\forall k,n,m\in \mathcal{K}\setminus \{k\}.
\end{align}
For the newly constructed constraint \eqref{PS36}, we provide the following three facts. Firstly, we remark that \eqref{PS36} is stricter than the original exponential cone constraint \eqref{PS16a}, which implies that the constraint \eqref{PS36} gives a safe approximation of the constraint \eqref{PS16a}.
Secondly, we note that \eqref{PS36} and \eqref{PS16a} are equivalent when $\theta_{k,m}[n]=\tilde{\theta}_{k,m}[n]$.
%Additionally, the proposed P-SCA iteration algorithm (shown at the next subsection) guarantees that $\theta_{k,m}[n]$ and $\tilde{\theta}_{k,m}[n]$ are very close at the convergence point.
Thirdly, although \eqref{PS36} cannot be represented as the SOC form directly,
we can introduce several slack variables to convert it into a SOC form.
In the following, we present a proposition to convert \eqref{PS36} into the SOC form.
\begin{proposition}\label{proposition2}
The constraint \eqref{PS36} can be represented as the following SOC constraints:
\begin{align}
&\left\|\frac{\pi_{k,m}[n]P^k_s[n]\lambda_{k,m}}{\sigma^2},\frac{\varsigma_m[n]-\tau_k[n]}{2}\right\|\leq \frac{\varsigma_m[n]\!+\!\tau_k[n]}{2},\label{SOC6}\\
&\Bigg\|\frac{\varpi_{k,m}[n]-\pi_{k,m}[n]-\log\left(\hat{\theta}_{k,m}[n]\right)-2}{2},\sqrt{2}\hat{\theta}_{k,m}^{\frac{1}{4}}[n]\Bigg\|\nonumber\\
&~~~~~~~~~\leq \frac{\varpi_{k,m}[n]+\pi_{k,m}[n]+\log\left(\hat{\theta}_{k,m}[n]\right)+2}{2},\label{SOC7}\\
%&-\log\left(1-\frac{1}{\tilde{\theta}_{k,m}[n]}\right)-2+\frac{2\sqrt{1-\frac{1}{\tilde{\theta}_{k,m}[n]}}}{\varpi_{k,m}[n]}\leq \pi_{k,m}[n],\label{PS40a}\\
&\left\|\varpi_{k,m}[n],\frac{-\xi_{k,m}[n]}{2}\right\|\leq \frac{2-\xi_{k,m}[n]}{2},\label{SOC8}\\
%&\sqrt{1-\xi_{k,m}[n]}\geq \varpi_{k,m}[n],\label{PS40b}\\
&\left\|\frac{\theta_{k,m}[n]-\xi_{k,m}[n]}{2},1\right\|\leq \frac{\theta_{k,m}[n]+\xi_{k,m}[n]}{2},\label{SOC9}
\end{align}
where $\pi_{k,m}[n]$, $\varpi_{k,m}[n]$ and $\xi_{k,m}[n]$, $\forall k,n,m\in \mathcal{K}\setminus \{k\}$, are introduced slack variables, while $\hat{\theta}_{k,m}[n]\triangleq 1-\frac{1}{\tilde{\theta}_{k,m}[n]}$.
\end{proposition}
\begin{IEEEproof}
The detailed proof is provided in Appendix \ref{App_B}.
\end{IEEEproof}

So far, we have represented the exponential cone constraint \eqref{PS16a} into the SOC constraints \eqref{SOC6}, \eqref{SOC7}, \eqref{SOC8} and \eqref{SOC9}.
Following the above transformations, problem ($\mathbf{P2.3}$) can be approximately rewritten as the SOCP, which is given by

\begin{align}\label{PF2_4}
&(\mathbf{P2.4}):\max_{\substack{\mathbf{Q},\mathbf{A}, \eta,\mathbf{P}_{\mathrm{U}},\mathbf{R}_{\mathrm{E}},\phi\\
\mathbf{U},\mathbf{V},\mathbf{\Theta},\mathbf{S}_1,\mathbf{S}_2,\mathbf{S}_3,\mathbf{S}_4,\mathbf{S}_5,\mathbf{S}_6}}~\eta-\omega\phi\nonumber\\
&\mathrm{s.t.}~\eqref{PF1c},\eqref{PF1e},\eqref{PF1f},\eqref{PF1g},\eqref{PS2b},\eqref{PS3},\eqref{PS14},\eqref{PS15_2},\nonumber\\
&~~~~~\eqref{PS22},\eqref{SOC2},\eqref{SOC3},\eqref{SOC4},\eqref{SOC5},\eqref{SOC6},\eqref{SOC7},\eqref{SOC8},\eqref{SOC9},
\end{align}
where $\mathbf{S}_3\triangleq\{\zeta[n],\forall n\}$, $\mathbf{S}_4\triangleq\{\pi_{k,m}[n],\forall k,n,m\in \mathcal{K}\setminus \{k\}\}$, $\mathbf{S}_5\triangleq\{\varpi_{k,m}[n],\forall k,n,m\in \mathcal{K}\setminus \{k\}\}$, and $\mathbf{S}_6\triangleq\{\xi_{k,m}[n],\forall k,n,m\in \mathcal{K}\setminus \{k\}\}$.

We note that the SOCP ($\mathbf{P2.4}$) introduces slack variables $\zeta[n]$, $\pi_{k,m}[n]$, $\varpi_{k,m}[n]$, and $\xi_{k,m}[n]$. However, the complexity of solving an SOCP is greatly reduced
compared to solving GNCP of a similar size directly due to the tremendous progress in current SOCP solvers \cite{Nguyen2019Energy}.

\subsection{Proposed P-SCA Algorithm for Solving ($\mathbf{P2}$)}

\begin{algorithm}[t]
\caption{P-SCA algorithm for Solving ($\mathbf{P2}$)}\label{alg1}
\begin{algorithmic}[1]
\STATE Given a feasible point $\mathcal{\tilde{Z}}^0$ and an initial penalty parameter $\omega^0$; Given $c>1$ and $\omega_{\max}$; Set $r=0$.
\REPEAT
\STATE {Solve ($\mathbf{P2.4}$) with given a feasible point $\mathcal{\tilde{Z}}^r$ and obtain the current optimal solution $\mathcal{Z}^{r+1}$.}
\STATE {Update $\omega^{r+1}=\min\{c\omega^r,\mu_{\max}\}$ and set $\mathcal{\tilde{Z}}^{r+1}=\mathcal{Z}^{r+1}$; Set the iteration number $r=r+1$.}
\UNTIL {Convergence.}
\end{algorithmic}
\end{algorithm}

We recall that the original mixed-integer non-convex optimization problem $(\mathbf{P2})$ is first transformed into $(\mathbf{P2.2})$, and two problems are equivalent when $\phi\rightarrow 0$. Then, $(\mathbf{P2.2})$ is converted into a convex problem $(\mathbf{P2.3})$, which is categorized as GNCP. To reduce the computational complexity of GNCP $(\mathbf{P2.3})$, we further transform it into a SOCP $(\mathbf{P2.4})$.
We note that the feasible set of $(\mathbf{P2.4})$ is stricter than that of $(\mathbf{P2.2})$ since a restrictive approximation method is adopted. As a result, any feasible solution to $(\mathbf{P2.4})$ is also feasible to $(\mathbf{P2.2})$.
In addition, problem $(\mathbf{P2.2})$ can be solved by solving $(\mathbf{P2.4})$ iteratively.
The detailed algorithm is shown in Algorithm~\ref{alg1}, where $\mathcal{\tilde{Z}}^r\triangleq\{\tilde{\alpha}_k^r[n],\mathbf{\tilde{q}}_u^r[n],\tilde{P}_u^r[n],$ $\tilde{R}_{k,e}^r[n],\tilde{\mu}_k^r[n],\tilde{\nu}_k^r[n],\tilde{\theta}_{k,m}^r[n],\tilde{\varsigma}_m^r[n],\tilde{\tau}_k^r[n]\}$,  $\mathcal{Z}^r\triangleq\{\alpha_k^r[n],$ $\mathbf{q}_u^r[n],P_u^r[n],R_{k,e}^r[n],\mu_k^r[n],\nu_k^r[n],\theta_{k,m}^r[n],\varsigma_m^r[n],\tau_k^r[n]\}$, and $r$ denotes the $r$-th iteration. We note that the penalty parameter $\omega$ determines the relaxation level of problem $(\mathbf{P2.2})$. A large value of $\omega$ will force $\phi=0$, leading to $\alpha_k[n]\in \{0,1\}$.
We also note that the initial penalty parameter is set to a small value to provide high relaxation for the scheduling variable $\alpha_k[n]$.
The penalty parameter $\omega$ is increased by a constant $c>1$ after each iteration until a upper bound $\omega_{\max}$ is achieved to guarantee that $\phi=0$. Furthermore, numerical results show that $\phi$ is eventually equal to $0$ when the proposed P-SCA algorithm converges, which further verifies that $(\mathbf{P2})$ and $(\mathbf{P2.2})$ are equivalent.

\section{numerical results}

In this section, we provide numerical results to evaluate the effectiveness of our developed P-SCA algorithm. To demonstrate the benefit of our developed joint optimization of the UAV's trajectory and AN transmit power as well as the redundancy rate and the SN scheduling strategy (denoted as J-TPRS scheme), we compare it with a FT-PRS scheme. Specifically, the FT-PRS scheme only
designs the AN transmit power and redundancy rate together with the SN scheduling strategy based on Algorithm~\ref{alg1}, while the UAV's trajectory is fixed with circular trajectory in which the circular trajectory is given in \cite{JointWu2018}.
%We note that both the JTP and STP schemes are based on our proposed P-SCA algorithm.
Unless stated otherwise, the system parameters are set as below. The number of SNs is set as $K=4$ and the corresponding horizontal coordinates of each SN on the ground are set as $[-240, -160]^T$, $[160, -160]^T$, $[240, 80]^T$ and $[0, 160]^T$. The UAV's maximum AN transmit power and each SN's transmit power are assumed to be $P_{\max}^u=36~\mathrm{dBm}$ and $P^k_s[n]=30~\mathrm{dBm}, \forall k,n$, respectively. The UAV's maximum flying speed and its flying altitude are set as $V_{\max}=10~\mathrm{m/s}$ and $H=100~\mathrm{m}$, respectively. The time slot length is $\delta_t=1~\mathrm{s}$. Other simulation parameters are set as: $\sigma_u^2=-110~\mathrm{dBm}$, $\sigma_m^2=-110~\mathrm{dBm}$, $\forall m\in \mathcal{K}$, $\lambda_{u,u}=-60~\mathrm{dB}$, $\rho=-60~\mathrm{dB}$, $\beta_0=-60~\mathrm{dB}$, $T=210~\mathrm{s}$, $\epsilon_r=0.05$ and $\epsilon_s=0.05$.
\begin{figure}[!t]
  \centering
  % Requires \usepackage{graphicx}
  \includegraphics[width=3.49in, height=2.8in]{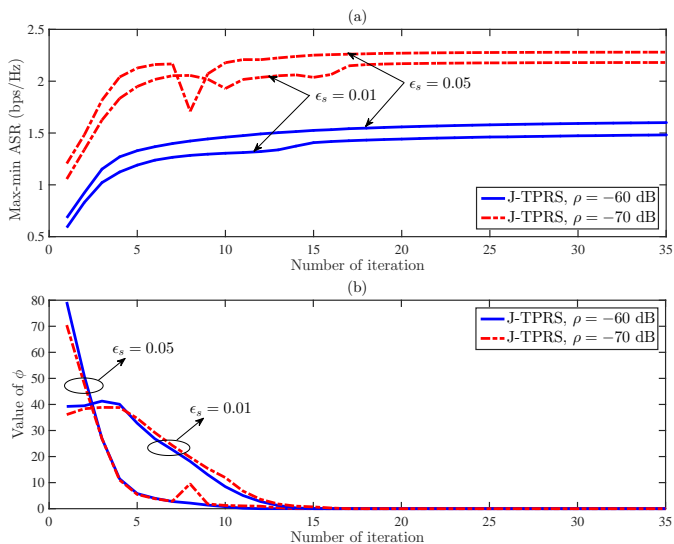}
  \caption{Convergence performance of the proposed Algorithm~\ref{alg1} for different values of $\rho$ and  $\epsilon_s$.}\label{convergence}
\end{figure}

In Fig.~\ref{convergence}, we examine the convergence behavior of the proposed P-SCA algorithm, where Fig.~\ref{convergence}(a) and Fig.~\ref{convergence}(b) show the variation of max-min ASR over iterations and the value of introduced slack variable $\phi$ over iterations, respectively. In Fig.~\ref{convergence}(a),
we first observe that the max-min ASR may be unstable at some intermediate iterations. This is due to the variation of penalty term $\omega\phi$ in the objective function. We also observe that the proposed algorithm converges within a few tens of iterations. This is due to the fact that Algorithm~\ref{alg1} is equivalent to the traditional SCA algorithm when $\omega$ reaches its upper bound  (i.e., the penalty parameter $\omega$ is increased after each iteration until reaching $\omega_{\max}$), while the convergence of the SCA algorithm has been proven in \cite{Li2013Coordinated}.
%we observe that the proposed algorithm converges within a few tens of iterations.
%We note that the max-min ASR may be unstable at some intermediate iterations due to the variation of the  penalty term $\omega\phi$ in objective function. Since the penalty parameter $\omega$ is increased at each iteration until reaching $\omega_{\max}$, Algorithm~\ref{alg1} can converges to a stationary point after penalty parameter $\omega$ is fixed.
In Fig.~\ref{convergence}(b), we observe that the value of $\phi$ converges to zero, which implies that the obtained scheduling variables $\alpha_k[n]$, $\forall k,n$, are binary and the outputs of Algorithm~\ref{alg1} are feasible solution to the original mixed-integer optimization problem $(\mathbf{P2})$.

\begin{figure}[!t]
  \centering
  % Requires \usepackage{graphicx}
  \includegraphics[width=3.49in, height=2.8in]{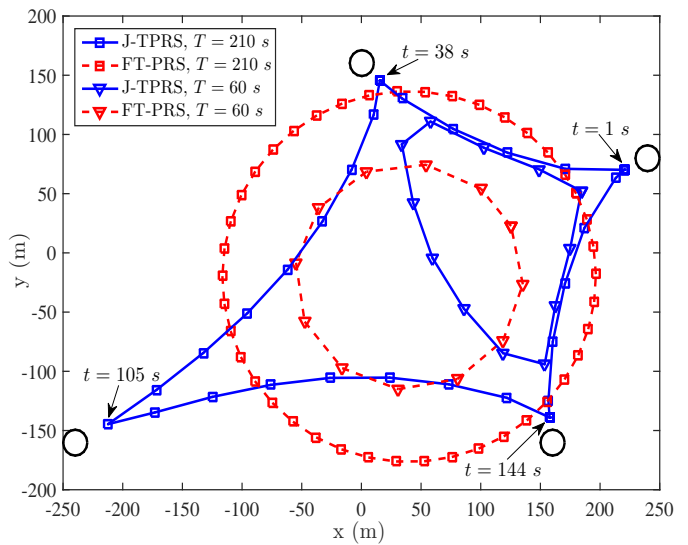}
  \caption{UAV's trajectories achieved by the J-TPRS and FT-PRS schemes for different values of the flight period $T$.}\label{Trajectory}
\end{figure}

\begin{figure}[!t]
  \centering
  % Requires \usepackage{graphicx}
  \includegraphics[width=3.49in, height=2.8in]{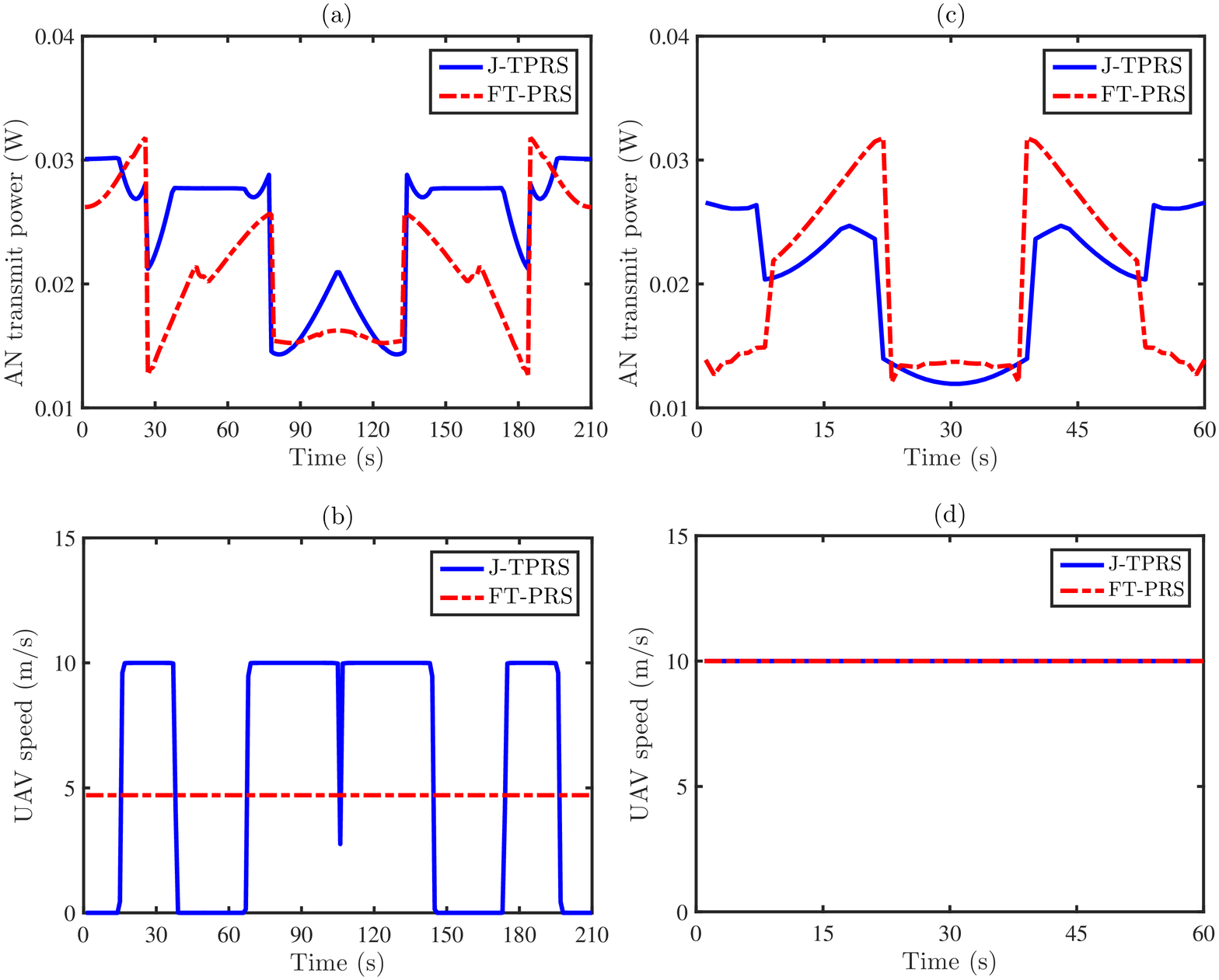}
  \caption{The UAV's AN transmit power and its flying speed for different values of the flight period $T$, where $T=210~\mathrm{s}$ for (a) and (b), while $T=60~\mathrm{s}$ for (c) and (d).}\label{Power_Speed}
\end{figure}

In Fig.~\ref{Trajectory}, we plot the trajectories of the UAV achieved by the J-TPRS scheme and the FT-PRS scheme  with different flight periods $T$, where the location of each SN is marked with $\bigcirc$. From this figure, we first observe that as flight period $T$ increases, the UAV takes full advantage of its controllable mobility to adaptively adjust its trajectory to move closer to each SN. For example, for $T=210~\mathrm{s}$, in the J-TPRS scheme the UAV always flies at the maximum speed when the UAV is between any two SNs, and it always reduces its flying speed or even hovers for a period of time when it arrives at a location near each SN, so that more confidential information can be transmitted from the scheduled SN over a better ground-to-air channel.
This phenomenon can be directly confirmed by the flight speed of the UAV, which is shown in Fig.~\ref{Power_Speed}(b). In Fig.~\ref{Trajectory}, we also observe that when $T$ is sufficient large (e.g., $T=210~\mathrm{s}$), the trajectory achieved by the J-TPRS scheme always shrinks inward to generate more interference to other unscheduled SNs. In addition, for $T=60~\mathrm{s}$, the UAV flies at the maximum speed within the limited flight period $T$
in order to get as close to each ground SN as possible for shorter LoS communication links.

%This observation can be confirmed by Fig.~\ref{Power_Speed}(d).

\begin{figure}[!t]
  \centering
  % Requires \usepackage{graphicx}
  \includegraphics[width=3.49in, height=2.8in]{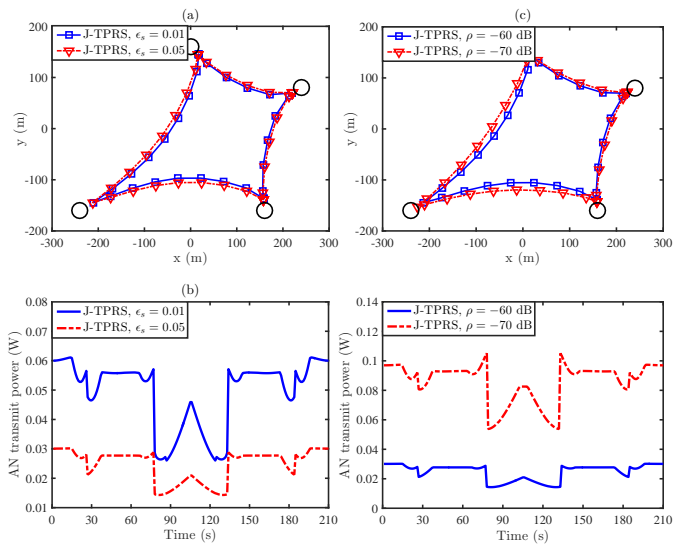}
  \caption{UAV's trajectories and the AN transmit power achieved by the J-TPRS scheme for different values of $\epsilon_s$ and $\rho$.}\label{Trajectory_Power}
\end{figure}
In Fig.~\ref{Power_Speed}, we plot the UAV's AN transmit power and its flying speed achieved by the J-TPRS and FT-PRS schemes for different values of the flight period $T$. In Fig.~\ref{Power_Speed}(a) and Fig.~\ref{Power_Speed}(c), we first observe that the UAV's AN transmit power is symmetrical, and the symmetry points are located at $105~\mathrm{s}$ and $30~\mathrm{s}$, respectively. This is due to the fact that the trajectories of the UAV achieved by the J-TPRS and FT-PRS schemes for $T=210~\mathrm{s}$ and $T=60~\mathrm{s}$ are symmetrical. From Fig.~\ref{Power_Speed}(a) and Fig.~\ref{Power_Speed}(c), we also observe that when the UAV reaches the area near the symmetrical point, the AN transmit power achieved by the J-TPRS scheme is obviously smaller than that of other locations. This is because the SN in the lower left corner is scheduled at this period of time and the SN is far away from other SNs. Thus, the UAV chooses a smaller AN transmit power to reduce self-interference while satisfying the required SOP constraint. In Fig.~\ref{Power_Speed}(a) and Fig.~\ref{Power_Speed}(b), as expected we observe that the AN transmit power achieved by the J-TPRS scheme remains constant when the UAV speed is zero. From Fig.~\ref{Power_Speed}(b) and Fig.~\ref{Power_Speed}(d), we observe that the UAV always flies with a constant speed in the FT-PRS scheme, which is consistent with the introduced circular trajectory.
Finally, we should point out that there is no exact relationship between the AN transmit power achieved by the J-TPRS and FT-PRS schemes. Intuitively, we may think that when $T$ is sufficiently large (e.g., $T=210~\mathrm{s}$), the AN transmit power of FT-PRS scheme is higher than that of J-TPRS scheme in most of the flight period $T$, since the circular trajectory always expands outwards. However, this contradicts to the observation made in Fig.~\ref{Power_Speed}(a), which is due to the fact that the values of $R_{k,e}[n]$ are different (i.e., $R_{k,e}[n]$ is an optimization variable and a larger value of $R_{k,e}[n]$ may require a smaller value of AN transmit power) in the J-TPRS and FT-PRS schemes.

\begin{figure}[!t]
  \centering
  % Requires \usepackage{graphicx}
  \includegraphics[width=3.49in, height=2.8in]{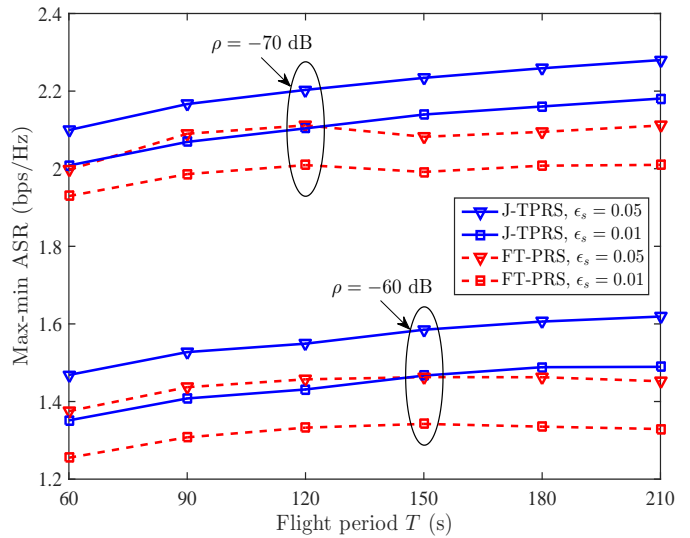}
  \caption{Max-min ASR (average secrecy rate) achieved by the J-TPRS and FT-PRS schemes versus flight period $T$ for different $\epsilon_s$ and $\rho$.}\label{ASR}
\end{figure}

In Fig.~\ref{Trajectory_Power}, we plot the UAV's trajectories and AN transmit power achieved by our developed J-TPRS scheme for different SOP levels $\epsilon_s$ and different self-interference levels $\rho$.
In Fig.~\ref{Trajectory_Power}(a) and Fig.~\ref{Trajectory_Power}(b), we first observe that as the $\epsilon_s$ decreases, the UAV's trajectory shrinks inward and its AN transmit power increases. This is due to the fact that the SOP constraint becomes stricter as $\epsilon_s$ decreases. Thus, the UAV selects a trajectory closer to each SN and uses a larger AN transmit power to satisfy the SOP constraint. In Fig.~\ref{Trajectory_Power}(c) and Fig.~\ref{Trajectory_Power}(d), we observe that the UAV's trajectory shrinks inward  as the self-interference level $\rho$ increases, and its AN transmit power decreases as $\rho$ increases. This is due to the fact that with the increasing of self-interference level $\rho$, the UAV prefers to select a trajectory closer to each SN to decrease its AN transmit power and reduce self-interference.

In Fig.~\ref{ASR}, we plot the max-min ASR achieved by the J-TPRS and FT-PRS schemes versus the flight period $T$ for different $\epsilon_s$ and $\rho$. In this figure, we first observe that the max-min ASR achieved by the J-TPRS scheme increases with flight period $T$. This is due to the fact that a larger $T$ provides a larger degree of freedom for UAV to adjust its flight trajectory to improve max-min ASR.
In this figure, we also observe that the max-min ASR obtained by the FT-PRS scheme does not increase with the increase of $T$ and the J-TPRS scheme always achieves a higher max-min ASR than the FT-PRS scheme. This observation demonstrates the importance of the UAV trajectory optimization and the advantage of the J-TPRS scheme.
In addition, as expected we observe that the max-min ASR obtained by both J-TPRS and FT-PRS schemes decreases as the self-interference level $\rho$ increases, while it increases as $\epsilon_s$ increases.

\section{Conclusions}
In this work, we addressed confidential data collection in UAV networks based on the physical layer security techniques.
We first derived analytical expressions for the ROP and SOP for the considered system, based on which we jointly optimized the UAV's trajectory and AN transmit power as well as the transmission rates and SN scheduling to maximize the minimum ASR subject to some specific constraints.
To tackle the formulated mixed-integer non-convex optimization problem, we first transformed it into a GNCP, and then we further converted the resultant optimization into a standard
SOCP to reduce the computational complexity. Finally, a novel iteration procedure based on P-SCA algorithm was developed to obtain a suboptimal solution to the formulated optimization problem.
Numerical results showed that the UAV's trajectory design is critical in the considered system and our developed solution achieves a significant performance gain relative to benchmark schemes.

\newcounter{mytempeqncnt2}
\begin{figure*}[tp]
\normalsize
\setcounter{mytempeqncnt2}{\value{equation}}
\setcounter{equation}{61}
\begin{align}\label{App_A4}
%&B_k^2[n]\triangleq\bigg(\left(\|\mathbf{\tilde{q}}_u[n]-\mathbf{w}_k\|^2+H^2\right)\left(-\rho \tilde{P}_u[n]\lambda_{u,u}\ln\epsilon_r+\sigma_u^2\right)+\beta_0 P^k_s[n]\bigg)\ln2,\\
%&B_k^3[n]\triangleq \frac{\beta_0 P^k_s[n]}{\|\mathbf{\tilde{q}}_u[n]-\mathbf{w}_k\|^2+H^2}, \\
%&B_k^4[n]\triangleq \frac{-\beta_0 P^k_s[n]\rho\lambda_{u,u}\ln\epsilon_r}{-\rho \tilde{P}_u[n]\lambda_{u,u}\ln\epsilon_r+\sigma_u^2},\\
&B_k[n]\triangleq \frac{-\beta_0 P^k_s[n]\rho\lambda_{u,u}\ln\epsilon_r(-P_u[n]+\tilde{P}_u[n])}{-\rho \tilde{P}_u[n]\lambda_{u,u}\ln\epsilon_r+\sigma_u^2}+\bigg(\left(\|\mathbf{\tilde{q}}_u[n]-\mathbf{w}_k\|^2+H^2\right)\left(-\rho \tilde{P}_u[n]\lambda_{u,u}\ln\epsilon_r+\sigma_u^2\right)+\beta_0 P^k_s[n]\bigg)\ln2\nonumber\\
&~~~~~~~\times\left[\log_2\left(1+\frac{\frac{\beta_0 P^k_s[n]}{\|\mathbf{\tilde{q}}_u[n]-\mathbf{w}_k\|^2+H^2}}{-\rho \tilde{P}_u[n]\lambda_{u,u}\ln\epsilon_r+\sigma_u^2}\right)-R_{k,e}[n]-\mu_k[n]\right]+\frac{\beta_0 P^k_s[n]\|\mathbf{\tilde{q}}_u[n]-\mathbf{w}_k\|^2}{\|\mathbf{\tilde{q}}_u[n]-\mathbf{w}_k\|^2+H^2}.
\end{align}
\setcounter{equation}{\value{mytempeqncnt2}}
\hrulefill
\vspace*{4pt}
\end{figure*}

\appendices
%========================================================================================

\section{Proof of Lemma \ref{lemma1}}\label{App_0}

We first substitute the $C_{k,u}[n]$ detailed in \eqref{Cap_u} into the definition of $p_{k}^{ro}[n]$ detailed in \eqref{ROP} and obtain that
\begin{align}\label{A_RO1}
&p_{k}^{ro}[n]=\mathrm{Pr}\left( |g_{u,u}[n]|^2>\frac{P^k_s[n]|h_{k,u}[n]|^2}{(2^{R_{k,u}[n]}-1)\rho P_u[n]}-\frac{\sigma_u^2}{\rho P_u[n]}\right).
\end{align}
We recall that the term $|g_{u,u}[n]|^2$ in \eqref{A_RO1} follows an exponential distribution
with parameter $\frac{1}{\lambda_{u,u}}$. Thus,
$p_k^{ro}[n]$ can be derived as
\begin{align}\label{A_RO2}
&p_{k}^{ro}[n]=\exp\left[\frac{-1}{\rho P_u[n]\lambda_{u,u}}\left(\frac{P^k_s[n]|h_{k,u}[n]|^2}{2^{R_{k,u}[n]}-1}-\sigma_u^2\right)\right]\nonumber\\
&=\exp\left[\frac{-1}{\rho P_u[n]\lambda_{u,u}}\left(\frac{\frac{\beta_0 P^k_s[n]}{\|\mathbf{q}_u[n]-\mathbf{w}_k\|^2+H^2} }{2^{R_{k,u}[n]}-1}-\sigma_u^2\right)\right].
\end{align}

Similarly, we substitute \eqref{Cap_m} into the definition of $p_{k}^{so}[n]$ detailed in \eqref{SOP} and rearrange \eqref{SOP} as
\begin{align}\label{A_SOP1}
&p_{k}^{so}[n]=\mathrm{Pr}\left(\max_{m\in \mathcal{K}\setminus \{k\}} \varrho_{u,k,m}[n]|g_{k,m}[n]|^2>2^{R_{k,e}[n]}-1\right),
\end{align}
where $\varrho_{u,k,m}[n]\triangleq\frac{P^k_s[n]}{ P_u[n]|h_{u,m}[n]|^2+\sigma_m^2}$. We note that the random variable $\varrho_{u,k,m}[n]|g_{k,m}[n]|^2$ involved in \eqref{A_SOP1} follows an exponential distribution
with parameter $\frac{1}{\varrho_{u,k,m}[n]\lambda_{k,m}}$. As such, $p_{k}^{so}[n]$ can be derived as
\begin{align}\label{A_SOP2}
&p_{k}^{so}[n]\!=\!1\!-\!\mathrm{Pr}\left(\max_{m\in \mathcal{K}\setminus \{k\}} \varrho_{u,k,m}[n]|g_{k,m}[n]|^2\leq2^{R_{k,e}[n]}\!-\!1\right)\nonumber\\
&=1-\prod_{m\in \mathcal{K}\setminus \{k\}}\left[1-\exp\left(\frac{1-2^{R_{k,e}[n]}}{\varrho_{u,k,m}[n]\lambda_{k,m}}\right)\right]\nonumber\\
&=1\!-\!\prod_{\!m\in \mathcal{K}\setminus \{k\}\!}\left[1\!-\!\exp\left(\frac{\frac{\beta_0 P_u[n]}{\|\mathbf{q}_u[n]-\mathbf{w}_m\|^2+H^2}+\sigma_m^2}{\frac{-P^k_s[n]\lambda_{k,m}}{2^{R_{k,e}[n]}-1}}\right)\right].
\end{align}
This completes the proof of Lemma~\ref{lemma1}.

\section{Proof of Proposition \ref{proposition1}}\label{App_A}
\subsubsection{The constraint \eqref{PS7}}
To rewrite the constraint \eqref{PS7} into the SOC form, let us first rearrange it as
\begin{align}\label{App_A1}
&\sum_{n=1}^N \big[(\alpha_k[n]-\mu_k[n])^2\big]\leq A_k[n],\forall k,
\end{align}
where
\begin{align}\label{App_A2}
A_k[n]\triangleq& \sum_{n=1}^N \big[2(\tilde{\alpha}_k[n]+\tilde{\mu}_k[n])(\alpha_k[n]+\mu_k[n])\nonumber\\
&-(\tilde{\alpha}_k[n]+\tilde{\mu}_k[n])^2\big]-4N\eta.
\end{align}
We observe that the LHS and the RHS of the \eqref{App_A1} are quadratic term and linear term, respectively.
Following the fact that linear term $A_k[n]$ defined in \eqref{App_A2} can be equivalently rewritten as $\left(\frac{A_k[n]+1}{2}\right)^2-\left(\frac{A_k[n]-1}{2}\right)^2$. Then, we can obtain the desired result in \eqref{SOC1}.
%\begin{align}\label{PS25}
%&\left\|\alpha_k[1]-\mu_k[1],\cdots, \alpha_k[N]-\mu_k[N], \frac{A_k[n]-1}{2}\right\|
%\nonumber\\
%&~~~~~~~~~~~~~~~~~~~~~~~~~~~~~~~~~~~~\leq \frac{A_k[n]+1}{2},\forall k.
%\end{align}

\subsubsection{The constraint \eqref{PS10}}
We note that \eqref{PS10} can be rearranged as
\begin{align}\label{App_A3}
&\frac{\beta_0 P^k_s[n]\|\mathbf{q}_u[n]-\mathbf{w}_k\|^2}{\|\mathbf{\tilde{q}}_u[n]-\mathbf{w}_k\|^2+H^2}
\leq B_k[n],\forall k,n,
\end{align}
where $B_k[n]$ is defined in \eqref{App_A4}, shown at the top of this page. Similar to the derivation of \eqref{App_A1}, the constraint \eqref{App_A3} can be rewritten as the standard SOC constraint shown in \eqref{SOC2}.
\setcounter{equation}{62}
%
% \eqref{App_A3} can be rewritten as the standard SOC constraint along with the same lines as the derivation of \eqref{PS7}, i.e.,
%$B_k^1[n]\triangleq\log_2\left(1+\frac{\frac{\beta_0 P^k_s[n]}{\|\mathbf{\tilde{q}}_u[n]-\mathbf{w}_k\|^2+H^2}}{-\rho \tilde{P}_u[n]\lambda_{u,u}\ln\epsilon_r+\sigma_u^2}\right)$,
%\begin{align}\label{PS27}
%%&B_k^2[n]\triangleq\bigg(\left(\|\mathbf{\tilde{q}}_u[n]-\mathbf{w}_k\|^2+H^2\right)\left(-\rho \tilde{P}_u[n]\lambda_{u,u}\ln\epsilon_r+\sigma_u^2\right)+\beta_0 P^k_s[n]\bigg)\ln2,\\
%%&B_k^3[n]\triangleq \frac{\beta_0 P^k_s[n]}{\|\mathbf{\tilde{q}}_u[n]-\mathbf{w}_k\|^2+H^2}, \\
%%&B_k^4[n]\triangleq \frac{-\beta_0 P^k_s[n]\rho\lambda_{u,u}\ln\epsilon_r}{-\rho \tilde{P}_u[n]\lambda_{u,u}\ln\epsilon_r+\sigma_u^2},\\
%&B_k[n]\triangleq \frac{-\beta_0 P^k_s[n]\rho\lambda_{u,u}\ln\epsilon_r(-P_u[n]+\tilde{P}_u[n])}{-\rho \tilde{P}_u[n]\lambda_{u,u}\ln\epsilon_r+\sigma_u^2}+\frac{\beta_0 P^k_s[n]\|\mathbf{\tilde{q}}_u[n]-\mathbf{w}_k\|^2}{\|\mathbf{\tilde{q}}_u[n]-\mathbf{w}_k\|^2+H^2}\nonumber\\
%&+\bigg(\left(\|\mathbf{\tilde{q}}_u[n]-\mathbf{w}_k\|^2+H^2\right)\left(-\rho \tilde{P}_u[n]\lambda_{u,u}\ln\epsilon_r+\sigma_u^2\right)+\beta_0 P^k_s[n]\bigg)\ln2\nonumber\\
%&\left[\log_2\left(1+\frac{\frac{\beta_0 P^k_s[n]}{\|\mathbf{\tilde{q}}_u[n]-\mathbf{w}_k\|^2+H^2}}{-\rho \tilde{P}_u[n]\lambda_{u,u}\ln\epsilon_r+\sigma_u^2}\right)-R_{k,e}[n]-\mu_k[n]\right].
%\end{align}

\subsubsection{The constraint \eqref{PS14}}
To transform \eqref{PS14} into an SOC constraint, we first rearrange it as
\begin{align}\label{App_A5}
&\sum_{k=1}^K \big[(\alpha_k[n]\!+\!\nu_k[n])^2\!+\!(\tilde{\alpha}_k[n]\!-\!\tilde{\nu}_k[n])^2\big]\leq C_k[n],\forall n,
\end{align}
where
\begin{align}\label{App_A6}
C_k[n]\triangleq4\epsilon_s+\sum_{k=1}^K2(\tilde{\alpha}_k[n]-\tilde{\nu}_k[n])(\alpha_k[n]-\nu_k[n]).
\end{align}
Then we can obtain the SOC form of the constraint \eqref{PS14} shown in \eqref{SOC3}.
%\begin{align}\label{PS29}
%&\bigg\|\alpha_1[n]+\nu_1[n],\cdots,\alpha_K[n]\!+\!\nu_K[n],\tilde{\alpha}_1[n]-\tilde{\nu}_1[n],\cdots,\nonumber\\
%&~~~~~~~~\tilde{\alpha}_K[n]-\tilde{\nu}_K[n],\frac{C_k[n]-1}{2}\bigg\|\leq \frac{C_k[n]+1}{2},\forall n.
%\end{align}
\subsubsection{The constraint \eqref{PS21}}
To deal with the constraint \eqref{PS21}, we introduce slack variables $\zeta[n],\forall n$, and equivalently rewrite it as
\begin{subequations}\label{App_A7}
\begin{align}
\|\mathbf{q}_u[n]-\mathbf{w}_m\|^2&\leq D_m[n],\forall n,m\in \mathcal{K}\setminus \{k\},\label{App_A7a}\\
\frac{1}{P_u[n]}&\leq \zeta[n],\forall n,\label{App_A7b}
%\frac{\frac{\beta_0}{\sigma^2}}{\frac{1}{\tilde{P}_u[n]}(\|\mathbf{\tilde{q}}_u[n]-\mathbf{w}_m\|^2+H^2)^2}
%
%
%\|\mathbf{q}_u[n]-\mathbf{w}_m\|^2\leq E_m[n]
%E_m[n]\triangleq
%&E_m^1[n]-E_m^2[n]\left(\zeta[n]-\frac{1}{\tilde{P}_u[n]}\right)-E_m^3[n]\left(\|\mathbf{q}_u[n]-\mathbf{w}_m\|^2-\|\mathbf{\tilde{q}}_u[n]-\mathbf{w}_m\|^2\right)\geq 0,\\
%&\frac{1}{P_u[n]}\leq \zeta[n],\forall n,\\
%&E_m^1[n]\triangleq \frac{\frac{\beta_0}{\sigma^2}\tilde{P}_u[n]}{\|\mathbf{\tilde{q}}_u[n]-\mathbf{w}_m\|^2+H^2}+1-\sqrt{\tilde{\varsigma}_m[n]}-\frac{1}{2}(\tilde{\varsigma}_m[n])^{\frac{-1}{2}}(\varsigma_m[n]-\tilde{\varsigma}_m[n])\\
%&E_m^2[n]\triangleq \frac{\frac{\beta_0}{\sigma^2}}{\frac{1}{(\tilde{P}_u[n])^2}(\|\mathbf{\tilde{q}}_u[n]-\mathbf{w}_m\|^2+H^2)}\\
%&E_m^3[n]\triangleq \frac{\frac{\beta_0}{\sigma^2}}{\frac{1}{\tilde{P}_u[n]}(\|\mathbf{\tilde{q}}_u[n]-\mathbf{w}_m\|^2+H^2)^2}
%\forall n,m\in \mathcal{K}\setminus \{k\},
\end{align}
\end{subequations}
where
\begin{align}\label{App_A8}
&D_m[n]\triangleq (\|\mathbf{\tilde{q}}_u[n]-\mathbf{w}_m\|^2+H^2)(2-\tilde{P}_u[n]\zeta[n])\nonumber\\
&~~~+\bigg[1-\sqrt{\tilde{\varsigma}_m[n]}-\frac{1}{2}(\tilde{\varsigma}_m[n])^{\frac{-1}{2}}(\varsigma_m[n]-\tilde{\varsigma}_m[n])\bigg]\nonumber\\
&~~~\times\frac{(\|\mathbf{\tilde{q}}_u[n]-\mathbf{w}_m\|^2+H^2)^2}{\frac{\beta_0}{\sigma^2}\tilde{P}_u[n]}+\|\mathbf{\tilde{q}}_u[n]-\mathbf{w}_m\|^2.
\end{align}
We note that $D_m[n]$ is a linear function with respect to the slack variables $\varsigma_m[n]$ and $\zeta_m[n]$. Thus, it can be equivalently rewritten as $\left(\frac{D_m[n]+1}{2}\right)^2-\left(\frac{D_m[n]-1}{2}\right)^2$. Following this fact, \eqref{App_A7a} can be rewritten as the SOC constraint \eqref{SOC4}.
%the SOC form of the constraint \eqref{PS30a} is given by
%\begin{align}\label{PS32}
%\left\|(\mathbf{q}_u[n]-\mathbf{w}_m)^T,\frac{D_m[n]-1}{2}\right\|^2&\leq \frac{D_m[n]+1}{2},
%\end{align}
%$\forall n,m\in \mathcal{K}\setminus \{k\}$.
In addition, \eqref{App_A7b} can be reformulated as the SOC constraint \eqref{SOC5} directly.

%\begin{align}\label{PS33}
%\left\|\frac{P_u[n]-\zeta[n]}{2},1\right\|^2\leq \frac{P_u[n]+\zeta[n]}{2}.
%\end{align}
%This completes the proof of Proposition~\ref{proposition1}.

\section{Proof of Proposition \ref{proposition2}}\label{App_B}
To proceed, we introduce slack variables $\pi_{k,m}[n]$, $\forall k,n,m\in \mathcal{K}\setminus \{k\}$, and rewrite \eqref{PS36} as
\begin{subequations}\label{PS37}
\begin{align}
&\frac{\sigma^2}{P^k_s[n]\lambda_{k,m}}\sqrt{\varsigma_m[n]\tau_k[n]}\geq \pi_{k,m}[n],\label{PS37a}\\
&-\log\left(\hat{\theta}_{k,m}[n]\right)-2+\frac{2\sqrt{\hat{\theta}_{k,m}[n]}}{\sqrt{1-\frac{1}{\theta_{k,m}[n]}}}\leq \pi_{k,m}[n].\label{PS37b}
\end{align}
\end{subequations}
$\forall k,n,m\in \mathcal{K}\setminus \{k\}$, where $\hat{\theta}_{k,m}[n]\triangleq 1-\frac{1}{\tilde{\theta}_{k,m}[n]}$.
We observe that \eqref{PS37a} can be rewritten as the SOC constraint directly, which is shown in \eqref{SOC6}.
 In addition, \eqref{PS37b} can be rewritten as
\begin{subequations}\label{PS39}
\begin{align}
&-\log\left(\hat{\theta}_{k,m}[n]\right)-2+\frac{2\sqrt{\hat{\theta}_{k,m}[n]}}{\varpi_{k,m}[n]}\leq \pi_{k,m}[n],\label{PS39a}\\
&\sqrt{1-\xi_{k,m}[n]}\geq \varpi_{k,m}[n],\label{PS39b}\\
&\frac{1}{\theta_{k,m}[n]}\leq \xi_{k,m}[n],\label{PS39c}
\end{align}
\end{subequations}
$\forall k,n,m\in \mathcal{K}\setminus \{k\}$, where $\varpi_{k,m}[n]$ and $\xi_{k,m}[n]$ are introduced slack variables. We note that \eqref{PS39a}, \eqref{PS39b} and \eqref{PS39c} admit the SOC-representation, which are given by \eqref{SOC7}, \eqref{SOC8} and \eqref{SOC9}.
This completes the proof of Proposition~\ref{proposition2}.

\bibliographystyle{IEEEtran}
\bibliography{IEEEfull,UAVsecure}

% Generated by IEEEtran.bst, version: 1.13 (2008/09/30)
\begin{thebibliography}{10}
\providecommand{\url}[1]{#1}
\csname url@samestyle\endcsname
\providecommand{\newblock}{\relax}
\providecommand{\bibinfo}[2]{#2}
\providecommand{\BIBentrySTDinterwordspacing}{\spaceskip=0pt\relax}
\providecommand{\BIBentryALTinterwordstretchfactor}{4}
\providecommand{\BIBentryALTinterwordspacing}{\spaceskip=\fontdimen2\font plus
\BIBentryALTinterwordstretchfactor\fontdimen3\font minus
  \fontdimen4\font\relax}
\providecommand{\BIBforeignlanguage}[2]{{%
\expandafter\ifx\csname l@#1\endcsname\relax
\typeout{** WARNING: IEEEtran.bst: No hyphenation pattern has been}%
\typeout{** loaded for the language `#1'. Using the pattern for}%
\typeout{** the default language instead.}%
\else
\language=\csname l@#1\endcsname
\fi
#2}}
\providecommand{\BIBdecl}{\relax}
\BIBdecl

\bibitem{Zhou2019Covert}
X.~{Zhou}, S.~{Yan}, J.~{Hu}, J.~{Sun}, J.~{Li}, and F.~{Shu}, ``Joint
  optimization of a {UAV}'s trajectory and transmit power for covert
  communications,'' \emph{IEEE Trans. Signal Process.}, vol.~67, no.~16, pp.
  4276--4290, Aug. 2019.

\bibitem{Zeng2016Wireless}
Y.~Zeng, R.~Zhang, and T.~J. Lim, ``Wireless communications with unmanned
  aerial vehicles: opportunities and challenges,'' \emph{IEEE Commun. Mag.},
  vol.~54, no.~5, pp. 36--42, May 2016.

\bibitem{Zhaonan2019Joint}
N.~{Zhao}, X.~{Pang}, Z.~{Li}, Y.~{Chen}, F.~{Li}, Z.~{Ding}, and M.~{Alouini},
  ``Joint trajectory and precoding optimization for {UAV}-assisted {NOMA}
  networks,'' \emph{IEEE Trans. Commun.}, vol.~67, no.~5, pp. 3723--3735, May
  2019.

\bibitem{JointWu2018}
Q.~Wu, Y.~Zeng, and R.~Zhang, ``Joint trajectory and communication design for
  multi-{UAV} enabled wireless networks,'' \emph{IEEE Trans. Wireless Commun.},
  vol.~17, no.~3, pp. 2109--2121, Mar. 2018.

\bibitem{UnmanHWANG2018}
H.~Wang, G.~Ren, J.~Chen, G.~Ding, and Y.~Yang, ``Unmanned aerial vehicle-aided
  communications: Joint transmit power and trajectory optimization,''
  \emph{IEEE Wireless Commu. Lett.}, vol.~7, no.~4, pp. 522--525, Aug. 2018.

\bibitem{Zhang2018Joint}
S.~Zhang, H.~Zhang, Q.~He, K.~Bian, and L.~Song, ``Joint trajectory and power
  optimization for {UAV} relay networks,'' \emph{IEEE Commun. Lett.}, vol.~22,
  no.~1, pp. 161--164, Jan. 2018.

\bibitem{UAV2018Xiao}
L.~Xiao, X.~Lu, D.~Xu, Y.~Tang, L.~Wang, and W.~Zhuang, ``{UAV} relay in vanets
  against smart jamming with reinforcement learning,'' \emph{IEEE Trans. Veh.
  Technol.}, vol.~67, no.~5, pp. 4087--4097, May 2018.

\bibitem{Zhan2018Energy}
C.~Zhan, Y.~Zeng, and R.~Zhang, ``Energy-efficient data collection in {UAV}
  enabled wireless sensor network,'' \emph{IEEE Wireless Commun. Lett.},
  vol.~7, no.~3, pp. 328--331, Jun. 2018.

\bibitem{Lyu2016Cyclical}
J.~Lyu, Y.~Zeng, and R.~Zhang, ``Cyclical multiple access in {UAV}-aided
  communications: A throughput-delay tradeoff,'' \emph{IEEE Wireless Commun.
  Lett.}, vol.~5, no.~6, pp. 600--603, Dec. 2016.

\bibitem{Chen2017Survey}
X.~{Chen}, D.~W.~K. {Ng}, W.~H. {Gerstacker}, and H.~{Chen}, ``A survey on
  multiple-antenna techniques for physical layer security,'' \emph{IEEE Commun.
  Surveys Tuts.}, vol.~19, no.~2, pp. 1027--1053, 2rd~Quart. 2017.

\bibitem{Chen2018Exploiting}
X.~{Chen}, Z.~{Zhang}, C.~{Zhong}, D.~W.~K. {Ng}, and R.~{Jia}, ``Exploiting
  inter-user interference for secure massive non-orthogonal multiple access,''
  \emph{IEEE J. Sel. Areas in Commun.}, vol.~36, no.~4, pp. 788--801, Apr.
  2018.

\bibitem{Shu2016Robust}
F.~{Shu}, X.~{Wu}, J.~{Li}, R.~{Chen}, and B.~{Vucetic}, ``Robust synthesis
  scheme for secure multi-beam directional modulation in broadcasting
  systems,'' \emph{IEEE Access}, vol.~4, pp. 6614--6623, 2016.

\bibitem{Chen2019Physical}
R.~{Chen}, C.~{Li}, S.~{Yan}, R.~{Malaney}, and J.~{Yuan}, ``Physical layer
  security for ultra-reliable and low-latency communications,'' \emph{IEEE
  Wireless Commun.}, vol.~26, no.~5, pp. 6--11, Oct. 2019.

\bibitem{Wu2019Safeguarding}
Q.~{Wu}, W.~{Mei}, and R.~{Zhang}, ``Safeguarding wireless network with {UAVs}:
  A physical layer security perspective,'' \emph{IEEE Wireless Commun.},
  vol.~26, no.~5, pp. 12--18, Oct. 2019.

\bibitem{Zhang2019Securing}
G.~{Zhang}, Q.~{Wu}, M.~{Cui}, and R.~{Zhang}, ``Securing {UAV} communications
  via joint trajectory and power control,'' \emph{IEEE Trans. Wireless
  Commun.}, vol.~18, no.~2, pp. 1376--1389, Feb. 2019.

\bibitem{Tang2019Secrecy}
J.~{Tang}, G.~{Chen}, and J.~P. {Coon}, ``Secrecy performance analysis of
  wireless communications in the presence of {UAV} jammer and randomly located
  {UAV} eavesdroppers,'' \emph{IEEE Trans. Inf. Forensics Security}, vol.~14,
  no.~11, pp. 3026--3041, Nov. 2019.

\bibitem{Zhou2018Improving}
Y.~{Zhou}, P.~L. {Yeoh}, H.~{Chen}, Y.~{Li}, R.~{Schober}, L.~{Zhuo}, and
  B.~{Vucetic}, ``Improving physical layer security via a {UAV} friendly jammer
  for unknown eavesdropper location,'' \emph{IEEE Trans. Veh. Technol.},
  vol.~67, no.~11, pp. 11\,280--11\,284, Nov. 2018.

\bibitem{Zhou2019UAV}
X.~{Zhou}, Q.~{Wu}, S.~{Yan}, F.~{Shu}, and J.~{Li}, ``{UAV}-enabled secure
  communications: Joint trajectory and transmit power optimization,''
  \emph{IEEE Trans. Veh. Technol.}, vol.~68, no.~4, pp. 4069--4073, Apr. 2019.

\bibitem{Cai2018Dual}
Y.~{Cai}, F.~{Cui}, Q.~{Shi}, M.~{Zhao}, and G.~Y. {Li}, ``Dual-{UAV}-enabled
  secure communications: Joint trajectory design and user scheduling,''
  \emph{IEEE J. Sel. Areas in Commun.}, vol.~36, no.~9, pp. 1972--1985, Sep.
  2018.

\bibitem{Cheng2019UAV}
F.~{Cheng}, G.~{Gui}, N.~{Zhao}, Y.~{Chen}, J.~{Tang}, and H.~{Sari},
  ``{UAV}-relaying-assisted secure transmission with caching,'' \emph{IEEE
  Trans. Commun.}, vol.~67, no.~5, pp. 3140--3153, May 2019.

\bibitem{Wang2017Improving}
Q.~Wang, Z.~Chen, W.~Mei, and J.~Fang, ``Improving physical layer security
  using {UAV}-enabled mobile relaying,'' \emph{IEEE Wireless Commun. Lett.},
  vol.~6, no.~3, pp. 310--313, Jun. 2017.

\bibitem{Lei2019Safeguarding}
H.~{Lei}, D.~{Wang}, K.~{Park}, I.~S. {Ansari}, J.~{Jiang}, G.~{Pan}, and
  M.~{Alouini}, ``Safeguarding {UAV} {IoT} communication systems against
  randomly located eavesdroppers,'' \emph{IEEE Internet Things J.}, to be
  published, 2019.

\bibitem{Liu2019Safeguarding}
C.~{Liu}, J.~{Lee}, and T.~Q.~S. {Quek}, ``Safeguarding {UAV} communications
  against full-duplex active eavesdropper,'' \emph{IEEE Trans. Wireless
  Commun.}, vol.~18, no.~6, pp. 2919--2931, Jun. 2019.

\bibitem{Cui2018Robust}
M.~{Cui}, G.~{Zhang}, Q.~{Wu}, and D.~W.~K. {Ng}, ``Robust trajectory and
  transmit power design for secure {UAV} communications,'' \emph{IEEE Trans.
  Veh. Technol.}, vol.~67, no.~9, pp. 9042--9046, Sep. 2018.

\bibitem{Gong2018Flight}
J.~{Gong}, T.~{Chang}, C.~{Shen}, and X.~{Chen}, ``Flight time minimization of
  {UAV} for data collection over wireless sensor networks,'' \emph{IEEE J. Sel.
  Areas Commun.}, vol.~36, no.~9, pp. 1942--1954, Sep. 2018.

\bibitem{Zhan2019Energy}
C.~{Zhan} and H.~{Lai}, ``Energy minimization in {Internet-of-Things} system
  based on rotary-wing {UAV},'' \emph{IEEE Wireless Commun. Lett.}, vol.~8,
  no.~5, pp. 1341--1344, Oct. 2019.

\bibitem{Zhou2011Rethinking}
X.~{Zhou}, M.~R. {McKay}, B.~{Maham}, and A.~{Hjorungnes}, ``Rethinking the
  secrecy outage formulation: A secure transmission design perspective,''
  \emph{IEEE Commun. Lett.}, vol.~15, no.~3, pp. 302--304, Mar. 2011.

\bibitem{Yan2018Three}
S.~{Yan}, N.~{Yang}, I.~{Land}, R.~{Malaney}, and J.~{Yuan}, ``Three
  artificial-noise-aided secure transmission schemes in wiretap channels,''
  \emph{IEEE Trans. Veh. Tech.}, vol.~67, no.~4, pp. 3669--3673, Apr. 2018.

\bibitem{Yan2016Artificial}
S.~{Yan}, X.~{Zhou}, N.~{Yang}, B.~{He}, and T.~D. {Abhayapala},
  ``Artificial-noise-aided secure transmission in wiretap channels with
  transmitter-side correlation,'' \emph{IEEE Trans. Wireless Commun.}, vol.~15,
  no.~12, pp. 8286--8297, Dec. 2016.

\bibitem{Yan2015Optimization}
S.~{Yan}, N.~{Yang}, G.~{Geraci}, R.~{Malaney}, and J.~{Yuan}, ``Optimization
  of code rates in {SISOME} wiretap channels,'' \emph{IEEE Trans. Wireless
  Commun.}, vol.~14, no.~11, pp. 6377--6388, Nov. 2015.

\bibitem{zhou2019twc}
X.~Zhou, S.~Yan, F.~Shu, R.~Chen, and J.~Li, ``{UAV}-enabled covert wireless
  data collection,'' \emph{arXiv:1906.08438}, Submitted, Jun. 2019.

\bibitem{Boyd}
S.~Boyd and L.~Vandenberghe, \emph{Convex Optimization}.\hskip 1em plus 0.5em
  minus 0.4em\relax Cambridge U.K.: Cambridge Univ. Press, 2004.

\bibitem{Tervo2015Optimal}
O.~{Tervo}, L.~{Tran}, and M.~{Juntti}, ``Optimal energy-efficient transmit
  beamforming for multi-user {MISO} downlink,'' \emph{IEEE Trans. Signal
  Process.}, vol.~63, no.~20, pp. 5574--5588, Oct. 2015.

\bibitem{Nguyen2019Energy}
K.~{Nguyen}, Q.~{Vu}, L.~{Tran}, and M.~{Juntti}, ``Energy efficiency fairness
  for multi-pair wireless-powered relaying systems,'' \emph{IEEE J. Sel. Areas
  in Commun.}, vol.~37, no.~2, pp. 357--373, Feb. 2019.

\bibitem{Li2013Coordinated}
W.~{Li}, T.~{Chang}, C.~{Lin}, and C.~{Chi}, ``Coordinated beamforming for
  multiuser {MISO} interference channel under rate outage constraints,''
  \emph{IEEE Trans. Signal Process.}, vol.~61, no.~5, pp. 1087--1103, Mar.
  2013.

\end{thebibliography}

% Note that IEEE does not put floats in the very first column - or typically
% anywhere on the first page for that matter. Also, in-text middle ("here")
% positioning is not used. Most IEEE journals use top floats exclusively.
% Note that, LaTeX2e, unlike IEEE journals, places footnotes above bottom
% floats. This can be corrected via the \fnbelowfloat command of the
% stfloats package.

\ifCLASSOPTIONcaptionsoff
  \newpage
\fi

%\begin{thebibliography}{1}

%XXXXXXXX

%\end{thebibliography}
%\clearpage
%\newpage

\end{document}